\documentclass[prb,reprint,aps,floatfix,showpacs,superscriptaddress,longbibliography]{revtex4-1}
\usepackage{amssymb}
\usepackage{bbm}
\usepackage{dsfont}

\usepackage{amsmath}
\usepackage{graphicx}
\usepackage[caption=false]{subfig}
\usepackage[colorlinks=true,linkcolor=blue,anchorcolor=red,citecolor=blue,urlcolor=blue]{hyperref}

\begin{document}
	\title{Quantum $z=2$ Lifshitz criticality in one-dimensional interacting fermions}
\author{Ke Wang}
\email{kewang07@uchicago.edu}
\affiliation{Kadanoff Center for Theoretical Physics,  James Franck Institute, Department of Physics, University of Chicago, Chicago, IL 60637, USA \\
}%
\date{\today}
\begin{abstract}
    We consider Lifshitz criticality (LC) with the dynamical critical exponent $z=2$ in one-dimensional interacting fermions with a filled Dirac Sea. We report that interactions have crucial effects on Lifshitz criticality. Single particle excitations are destabilized by interaction and decay into the particle-hole continuum, which is reflected in the logarithmic divergence in the imaginary part of one-loop self-energy. We show that the system is sensitive to the sign of interaction. Random-phase approximation (RPA) shows that the collective particle-hole excitations emerge only when the interaction is repulsive. The dispersion of collective modes is gapless and linear.
    If the interaction is attractive, the one-loop renormalization group (RG) shows that there may exist a stable RG fixed point described by two coupling constants. We also show that the on-site interaction (without any other perturbations at the UV scale) would always turn on the relevant velocity perturbation to the quadratic Lagrangian in the RG flow, driving the system flow to the conformal-invariant criticality.    In the numerical simulations of the lattice model at the half-filling, we find that, for either on-site positive or negative interactions, the dynamical critical exponent becomes $z=1$ in the infrared (IR) limit and the entanglement entropy is a logarithmic function of the system size $L$. The work paves the way to study one-dimensional interacting LCs.
    
\end{abstract}
\maketitle

{\it Introduction.} Strong infrared (IR) fluctuations
play an important role in low-dimensional field theories in the absence of a gap. In classical two-dimensional (2D) field theories with continuous symmetries, emerging logarithmic IR divergences
destroy the long-range order\cite{prl66Mermin,1970Berezinksy} and results in the absence of spontaneous symmetry breaking\cite{1973coleman,1973KT}. For quantum one-dimensional (1D) fermions, similar logarithmic divergences in loop diagrams destabilize single quasi-particles and lead to the breakdown of Fermi Liquids \cite{Abrikosov,Giamarchi}. The low-energy excitations become bosonic\cite{Tomonaga1950} upon a weak interaction turned on. This is the celebrated Luttinger Liquid (LL) theory\cite{Luttinger1963,Haldane1981,rmp12Glazman}. Non-Fermi liquids in two-dimension remain as a puzzle due to strong IR fluctuations\cite{PhysRevB.80.165102,https://doi.org/10.48550/arxiv.2208.00730,PhysRevB.82.075127}. 

Quantum criticality (QC) emerges when the phase transition happens between gapped phases. Many of them are described by conformal field theories\cite{pr69Wilson,npb84VBelavin,prl86Affleck,prl86Cardy,fran12conformal,math65Mattis} (CFTs). There are also occasions. One example is multi-critical points\cite{prl18Verresen,2021William,prr21Chepiga} with the non-unit dynamical critical exponent $z$. They are called Lifshitz criticalities (LCs)\cite{PhysRevLett.35.1678,prd09Per}. The non-relativistic exponent $z$ is not rare. For example, free Sch\"{o}dinger equation and unitary Fermi gas are characterzied by $z=2$\cite{PhysRevD.76.086004,Golkar2014}. Symmetry algebra is the Sch\"{o}dinger algebra\cite{PhysRevD.5.377,Niederer1973} instead of conformal algebra.

Interactions are essential to LCs since the interaction energy can exceed the kinetic energy. The scale of kinetic energy is $K\sim t/L^z$ for the 1D system with a finite-size $L$. The contact interaction has the scale $V \sim u_0/L$. Here $t$ and $u_0$ are two finite coefficients to ensure the energy dimension. The ratio of two energy scales is $V/K\sim (u_0/t) L^{z-1}$. Therefore, when $z>1$, the interaction effects dominate. 

In this paper, we consider one-dimensional interacting fermions with the dynamical critical exponent $z=2$ and a filled Fermi Sea. In this case, besides interaction energy exceeding kinetic energy, interaction also causes strong quantum fluctuations. For example, interactions cause the creation of virtual particle-hole pairs (vacuum polarizations) which renormalize the low-energy behaviors. Therefore, one has to consider interaction effects carefully to obtain the right low-energy descriptions. Here we study interacting LCs from both aspects of field theories and numerical simulations.    Below we start with the perturbative calculations. 

 
{\it Model and Perturbative theory.} The non-interacting $z=2$ LC emerges from a slightly generalized Su–Schrieffer–Heeger (SSH) model\cite{prl79Su},
\begin{eqnarray}
\label{1}
H=\sum_{i=1}^L \sum_{l=0,1,2} \Big( t_l \hat{c}^\dagger_{i-l,A} \hat{c}_{i,B}+h.c. \Big) + u_0 \hat{n}_{i,A} \hat{n}_{i,B}.
\end{eqnarray}
Here $\hat{c}^\dagger_{i,A/B}$ is two-species fermion operator, $L$ is the linear size of lattice and $\hat{n}_{i,A/B}=\hat c^\dagger_{i,A/B} \hat c_{i,A/B}$ is the particle number operator.  $t_l$ with $l=1,2,3$ are three hopping parameters and $ u_0$ is the interaction strength. 
The non-interacting
$z=2$ LC emerges at the half-filling of the lattice model with parameters $t_0=t_2=t_1/2$.
The low-energy effective theory of Eq.~\ref{1} around this point may be represented by the spectral basis,
\begin{eqnarray}
\label{2}
\hat{H}_{\text{low-energy}}=\int dx \sum_{s=\pm}s \hat\psi^\dagger_s(x) (-i\partial_x)^2  \hat \psi_s(x)\nonumber \\
+ u_0 \int dx \hat\psi^\dagger_+(x) \hat\psi^\dagger_-(x)\hat\psi_-(x) \hat\psi_+(x).
\end{eqnarray}
Here $\hat\psi^\dagger_s(x)$
is the field operator in the continuous space, $s=\pm$ are two spectral branches and $t_0$ (bandwidth) is set to be unit. The non-interacting Feynman propagator is given by
\begin{eqnarray}
\label{3}
G_{s}(k,\omega)=\frac{1}{\omega- sk^2+i \delta \cdot \text{sign}(s)},\,\,\, s=\pm.
\end{eqnarray}
Here $\text{sign}(s)$ is the sign of $s$, denoting the particle/hole excitation, and $\delta=0^+$. Since virtual particle-hole pairs renormalize quasi-particle and electronic interaction, we consider the polarization operator (PO). The definition is $\Pi_{s,s'}(q,\omega)=-i (2\pi)^{-2} \int  {dk}  {d\omega'} G_s(k,\omega') G_{s'}(k-q,\omega'-\omega)\nonumber$. After the integral, it is given by
\begin{eqnarray}
\label{PO}
\Pi_{s,s'}(q,\omega)= \frac{1}{4} \frac{|s-s'|}{\sqrt{ - 2s\omega +  q^2 }}.
\end{eqnarray}
There emerges an algebraic singularity $\Pi(q,0)\sim 1/q$, which is due to $z=2$ dynamical critical exponent. Poles of the polarization operator locate the minimum of the particle-hole continuum.   The energy of a particle-hole pair is characterized by two momenta, $\epsilon(k,q)=(k+q)^2+k^2$, which fills a continuum in the spectrum. The continuum has a lower bound with $q^2/2$ when $k=-q/2$. This is exactly the pole of Eq.~\ref{PO} when $s>0$. Notice that $\Pi(q,\omega )$ becomes imaginary when $s\omega>q^2/2$, indicating that quasi-particles decay into the particle-hole continuum.
 
\begin{figure}
		\centering
		\includegraphics[scale=0.2]{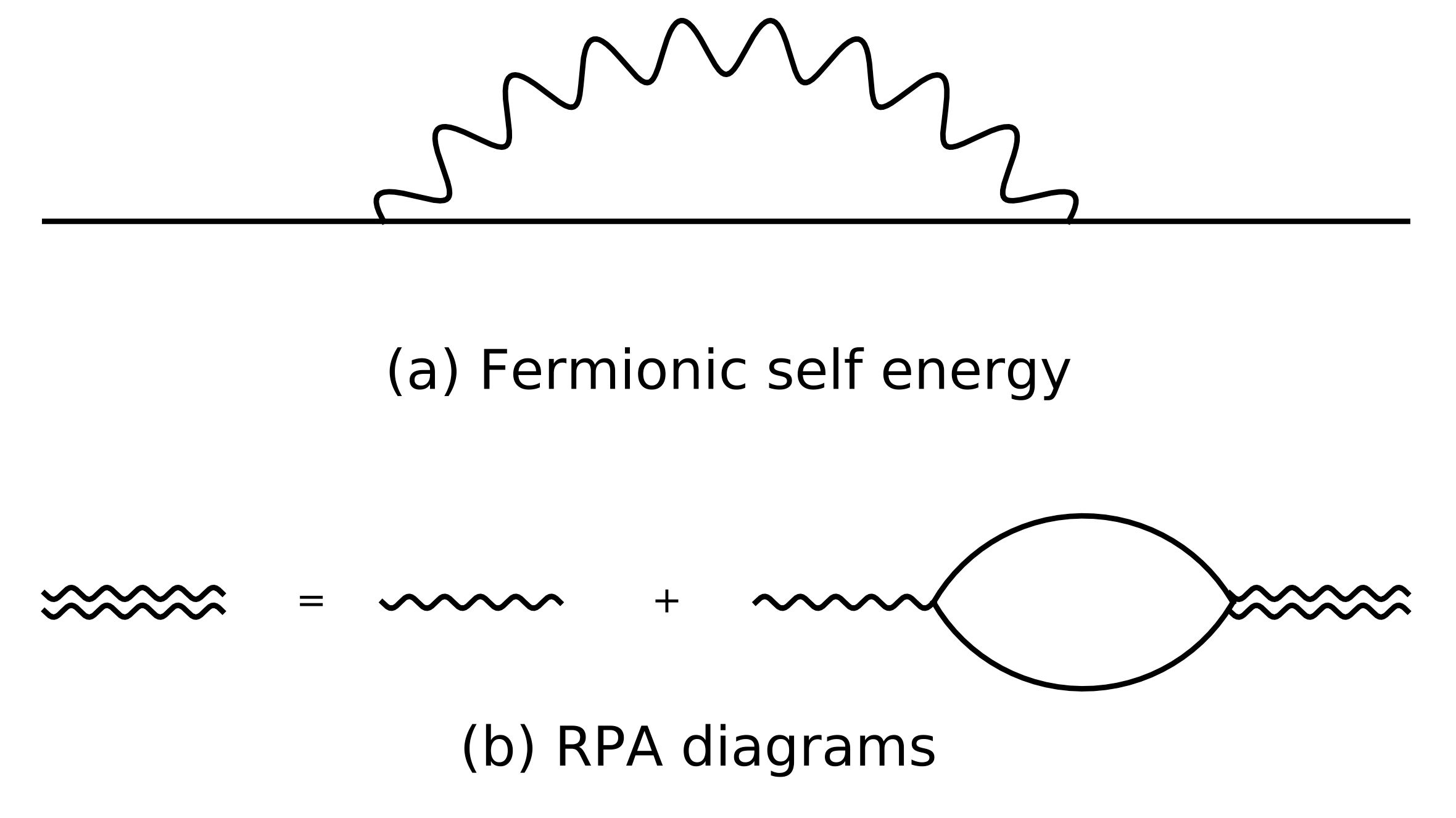} 
		\caption{
Two perturbative Feynman diagrams. (a) The fermionic self-energy. The solid line is the free fermion propagator and the wavy line is the electronic interaction. (b) RPA diagrams. The double wavy represents the interaction renormalized by loop diagrams.  }
		\label{diagram_1}
	\end{figure} 
 
 Inserting the one-loop diagram to Fig.~\ref{diagram_1}a, one may obtain the self-energy correction. For particle-like fermions, the correction reads $\Sigma_+(k,\omega)\simeq -i u_0^2 (2\pi)^{-2}\int  {d\nu} {dq }  G_{-}(k-q,\omega-\nu)    \Pi_{+-}(q,\nu)$. Performing the integral, we find the imaginary part of $\Sigma_+$,
\begin{eqnarray}
\label{5}
\text{Im} \Sigma_+(k,\omega)=-C u^2_0   \ln (\Lambda/|\omega-k^2|).
\end{eqnarray}
Here $\Lambda$ is the UV cut-off of the low-energy theory and $C$ is a positive constant. At the one-loop level, $C$ is given by $1/(2\pi)$. We check that the logarithmic divergence remains true at the RPA level\cite{supplement}. Therefore, one expects a vanishing lifetime of quasi-particles and unstable single-particle excitations. The zero lifetime has a very clear physical origin: single particle excitations are immersed in the particle-hole continuum and the interaction causes the single fermionic excitation to decay into the continuum. In the calculation, one may observe that the imaginary part of the polarization operator contributes a logarithmic term to Eq.~\ref{5}.

Nevertheless, collective excitations/modes still exist.
\begin{figure}
		\centering
		\includegraphics[scale=0.5]{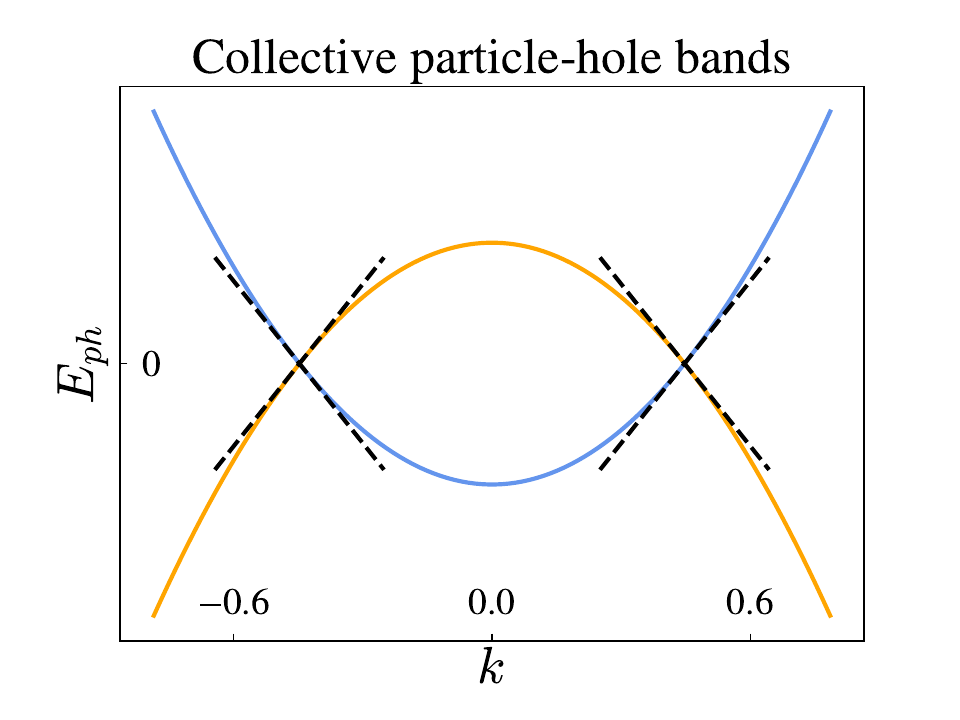} 
		\caption{ (Color Online) Energy of particle-hole collective modes $E_{ph}$. They are obtained from Eq.~\ref{6}. Two bands cross each other at two points in the momentum space, $k=\pm \sqrt{2 }u_0$. Near the crossing points, the dispersion becomes linear. Thus the low energy field theory is described by two copies of Luttinger liquids.     }
		\label{ph_C}
	\end{figure} 
The spectrum of particle-hole collective modes can be obtained by RPAs in Fig~\ref{diagram_1}b.
In LL theory, this method gives the same dispersion as the one obtained from the bosonization. RPA diagrams renormalize the bare Green function in Eq.~\ref{3} and the involved effective interaction is given by $u_s(q,\omega)\equiv u_0/(1-u_0\Pi_{s,-s}(q,\omega)),s=\pm$. The spectrum of particle-hole collective excitations can be extracted by the poles of the effective interaction $u_s(q,\omega)$. Poles only exist within repulsive interactions and collective excitations have the spectrum,
\begin{eqnarray}
\label{6}
E_{\text{s}}(q)=s(q^2/2-u^2_0/8),\quad s=\pm,
\end{eqnarray}
 when $u_0>0$. The spectrum of collective excitations is plotted in Fig.~\ref{ph_C}. Two bosonic bands intersect with each other at $q=u_0/2$. Thus the system at $u_0>0$ is gapless and the low energy physics is described as two copies of Luttinger Liquids, namely $c=2$. An additional quantitative prediction is that the velocity of the boson is linearly proportional to $u_0$. 

To make the perturbative description more complete, we also provide the real part of RPA self-energy, 
\begin{eqnarray}
\label{RPA}
\text{Re}\Sigma_{\text{RPA},+}=-\frac{u_0^2}{8} \text{sgn}(Q) -   \frac{u_0^3 d}{16\sqrt{u_0^2/4+2Q}},
\end{eqnarray}
Here $Q=\omega-k^2$, $d\equiv  1-\text{sgn}(u_0 Q)$, and we implicitly assume $u_0^2/4+2Q\geq 0$. When $u_0>0$, Eq.~\ref{RPA} has a singularity at $\omega-k^2=-u_0^2/8$, originating from the particle-hole collective modes.  When $u_0<0$, the $\text{Re}\Sigma_{\text{RPA},+}$ is always finite. The $\text{Re}\Sigma_{\text{RPA},+}$ varies from $u_0^2/8$ to $-u_0^2/8$ and cross zero at $\omega-k^2=3u_0^2/8$.  We would not consider the on-shell condition (solving $\omega-k^2-\text{Re}\Sigma_{\text{RPA},+}=0$), since the logarithmically large $\text{Im}\Sigma_{\text{RPA},+}$ invalidates the on-shell single-particle picture. 


{\it  Renormalization Group. } The bare Hamiltonian presented in Eq.~\ref{1} only assumes a constant $u$ at a given energy scale. To track the coupling $u$ at different energy scales, one may perform the momentum-shell renormalization of the bare action $S$. The method consists of iteratively lowering the energy scale $\Lambda$ and obtaining the effective action with a lower energy scale. At each step of renormalization, new scattering events may be generated in the effective action. This leads to the renormalization group equation (RGE) of the coupling constant $u$\cite{RevModPhys.66.129,PhysRevB.46.11749, PhysRevB.50.258,PhysRevA.75.033608,sachdev_2011}.
\begin{figure}
		\centering
		\includegraphics[scale=0.3]{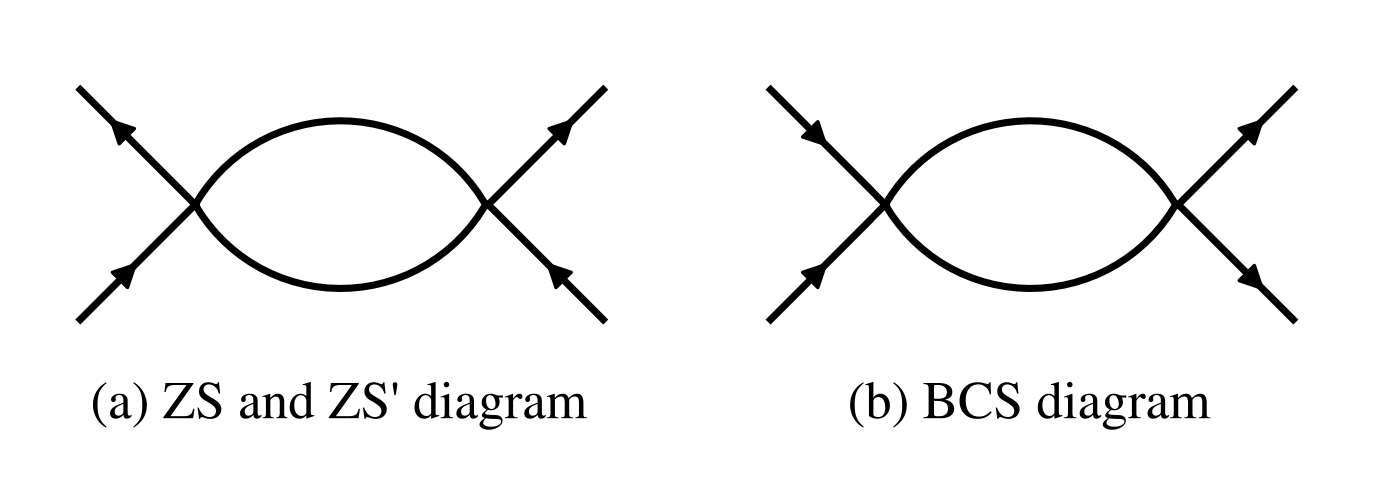} 
		\caption{ Three diagrams renormalize the coupling function $u(k_4,k_3;k_2,k_1)$. (a) ZS or ZS$'$ diagrams. The two diagrams share the same topology. But the momentum transfer at the vertex is different, $k_1-k_3$ for ZS and $k_1-k_4$ for ZS'. (b)  BCS diagram. Momentum transfer at the vertex is $k_1+k_2$.  }
		\label{potential}
	\end{figure}

 \begin{figure}[h]
		\centering
		\includegraphics[scale=0.4]{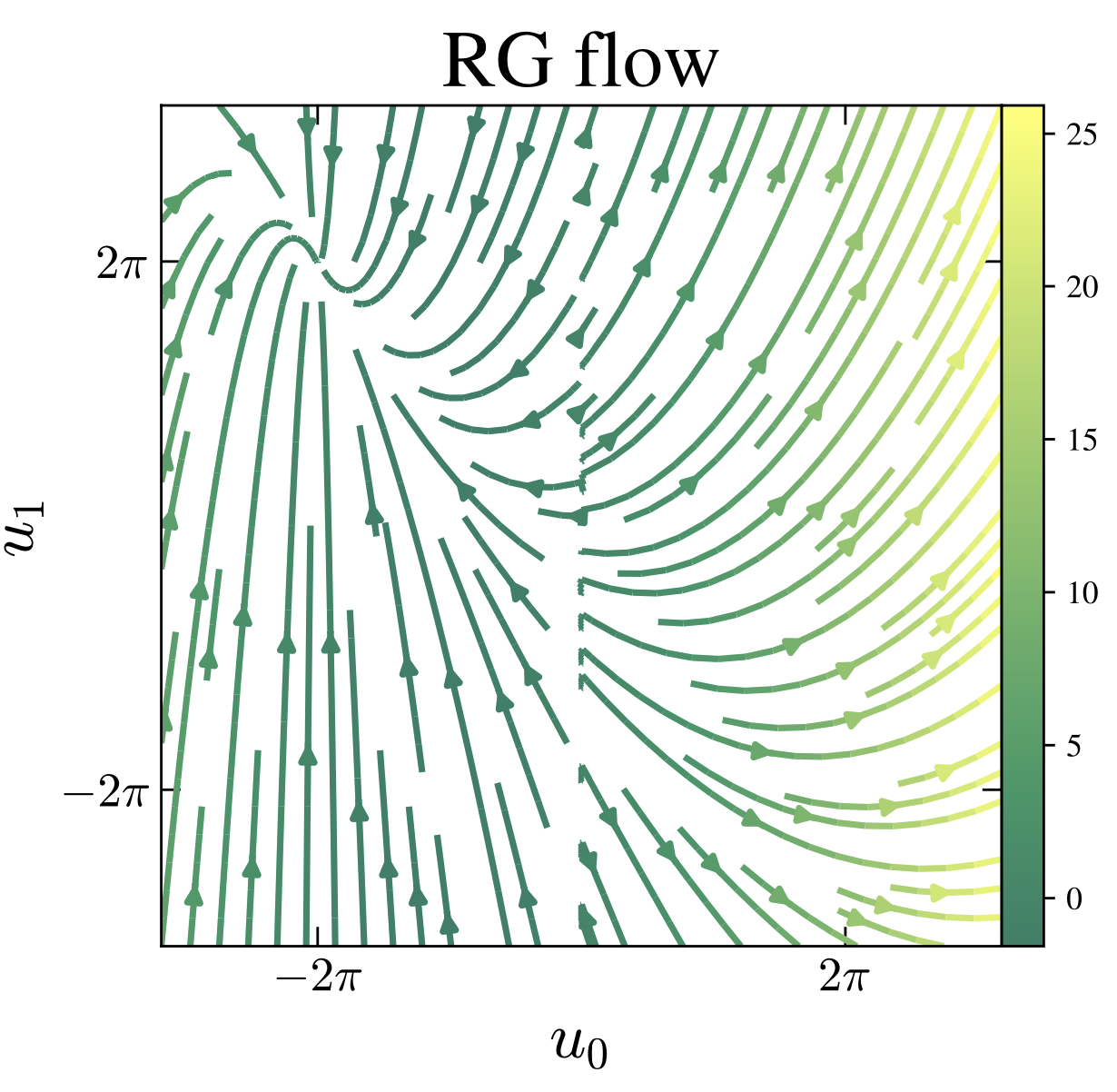} 
		\caption{ (Color Online)
	 RG flow of two coupling constants where $\Lambda$ is set to be unit. Colorbar is based on the magnitude of $\sqrt{(du_0/dl)^2+du_1/dl)^2}$. There exists one RG fixed point at $u_0/\Lambda=-u_1=-2\pi$.  }
		\label{run}
	\end{figure}

We start from the corresponding bare action in the Euclidean space-time given by $S=S_0+S_{\text{int}}$.
The non-interacting action reads
\begin{eqnarray} 
\label{7}
S_0&=& \sum_{s=\pm } \int \frac{dk d\omega}{(2\pi)^2}  \bar\psi_s(\omega k)(i\omega-s k^2)\psi_s(\omega k),
 \end{eqnarray}
and the interaction is written by a conventional symmetric potential, 
  \begin{eqnarray} 
S_{\text{int}}&=&\frac{1}{2!2!}  \prod_{i=1,2,3,4}
\int \frac{dk_id\omega_i}{(2\pi)^2}
 u(4321) 2\pi \delta_k\cdot   2\pi \delta_\omega \nonumber\\
&\times&    \bar\psi_{s_4}(\omega_4k_4)   \bar \psi_{s_3} (\omega_3k_3)  \psi_{s_2} (\omega_2k_2) \psi_{s_1}(\omega_1k_1).
 \end{eqnarray}
Here $\psi, \bar{\psi}$ is the Grassmannian variable, $u(4321)$ denotes the potential $u_{s_4s_3s_2s_1}(k_4k_3k_2k_1)$ maintaining momentum dependence and $\delta_k$ represents the momentum conservation $\delta(k_4+k_3-k_2-k_1)$.


At the tree-level RG (where $S_0$ is kept invariant), dimensions of momentum/frequency and the field operator are  given by 
\begin{eqnarray} 
  [k]=1, \, [\omega]=2,\, [\psi]=-5/2.
\end{eqnarray}
Therefore one has to consider two types of coupling constants in the symmetric potential. For example, one may expand one potential term $u_{+-+-}$,
\begin{eqnarray} 
u_{+-+-}(k_4k_3k_2k_1)\simeq u_{0}+ u_{1}(k_2-k_3) + \text{irrelevant}.
\end{eqnarray}
Any terms with frequency dependence and higher order ($\geq 2$) momentum dependence are irrelevant.
Other $u_{s_4s_3s_2s_1}$ can be obtained by exchanging indices of $u_{+-+-}$ from the symmetric property.  Note that terms linear in $k$ can be non-zero since particle/hole D. O. F. exists in the low-energy sector. One may find that dimension of $[u_0]=1$ and $[u_1]=0$. Therefore, $u_0$ is a {\it relevant} perturbation and $u_1$ is {\it marginal} perturbation. Generally, there are also possible relevant perturbations in the quadratic Lagrangian, $\sim  (m+vk) \bar\psi \psi .$
Here $m$ is the mass and $v$ is the velocity with the dimension $[m]=2$ and $[v]=1$. The quadratic relevant perturbation could be canceled by adding the counter terms by hand.


Now let us turn on $u_0$-interaction at the UV scale $\Lambda$ and integrate the momentum shell to see RG flow. At the one-loop level (second-order approximation of interactions), three types of diagrams are important, plotted in Fig.~\ref{potential} and called zero sound (ZS), zero sound$'$ (ZS$'$), and BCS diagrams. The ZS/ZS$'$ involves particle-hole channels while the BCS diagram is from the particle-particle channel. For the zero-temperature phase of $z=2$ LC, scatterings via particle-particle channels are absent. Therefore the ZS/ZS' diagrams are the main contribution. Performing one-loop integrals, one can obtain the following equations,
\begin{eqnarray}
\label{9}
\frac{du_0}{dl}&=&u_0 + \frac{u_0^2 }{2\pi\Lambda }, \nonumber
\\
\frac{du_1}{dl}&=&\frac{u_0^2 }{ \pi \Lambda^2 }+\frac{ u_0u_1 }{\pi \Lambda },
\end{eqnarray}
together with the $dv/dl=v+u_1 \Lambda/\pi$.
Here $l$ is the parameter running from $l=0$ (initial condition) to $l=+\infty$ (IR limit). Supposing that we always add some counter term to $v$ by hand, we can only focus on the RG flow of ${u_0,u_1}$ described by Eq.~\ref{9}. There exists a stable RG fixed point at $u_0=-2\pi \Lambda$ and $u_1=2\pi$. RG flow is plotted in Fig.~\ref{run}. If we consider the case like Eq.~\ref{1} with only one perturbation at UV, the $u_0$ turn on $v$ in RG, and $v$ flows to $+\infty$. This indicates that the IR theory of Eq.~\ref{1} is always described by the conformal invariant criticality, once $u_0\neq 0$.

\begin{figure} 
		\centering
		\includegraphics[scale=0.4]{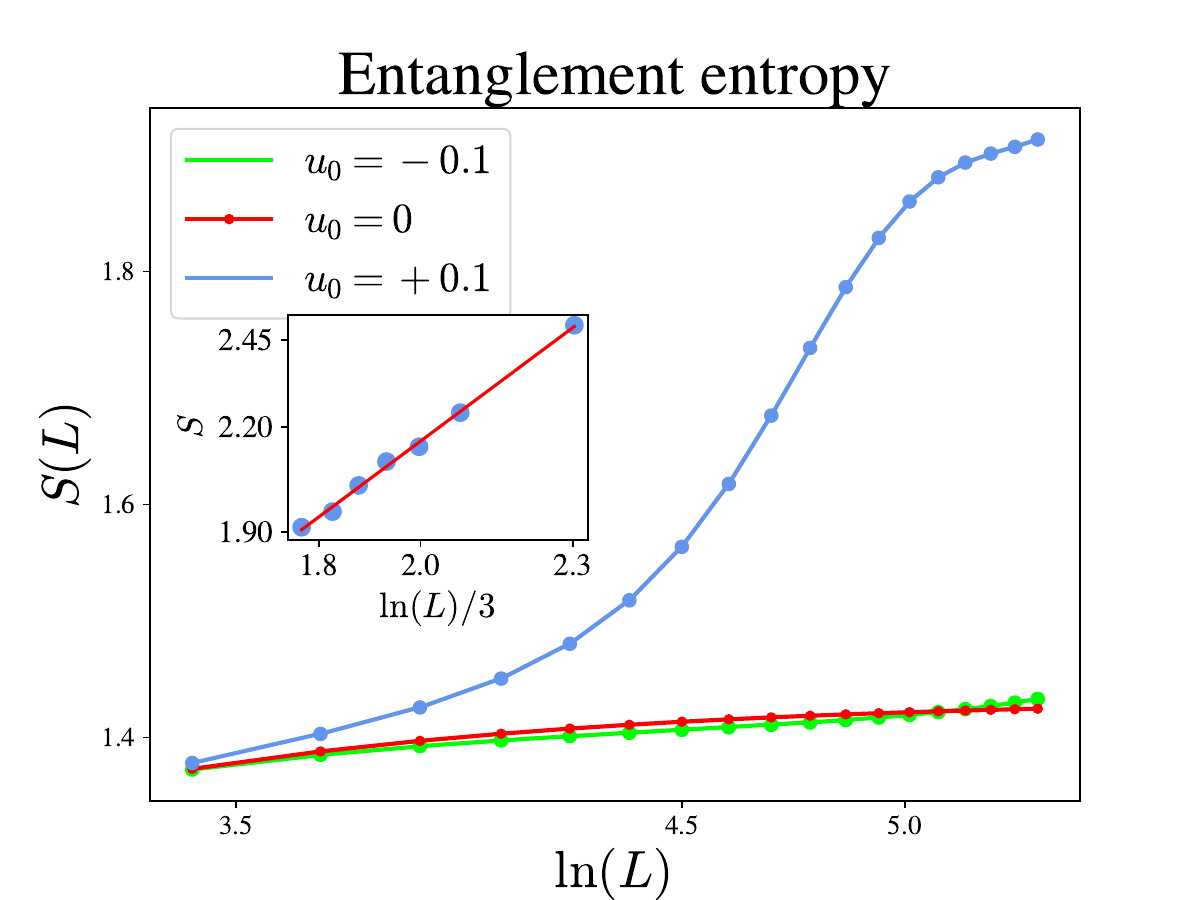} 
		\caption{ (Color Online)
  Results of Entanglement entropy $S(L)$ at $u_0=\pm 0.1$. Length varies from $L=30$ to $200$. The $S(L)$ exhibits sharply different behaviors for positive/negative couplings. Entropy climbs quickly at $u_0>0$ and deviates from the behavior of free theory when $L$ exceeds $30$. The entropy of $u_0=-0.1$ climbs much slower and only slightly deviate from the free theory when $L$ approaches $200$. The inset plots the entropy at $u_0=0.1$ with $L\geq 200$. The linear fit (red line) is performed for entropy versus $1/3 \ln L$ when $200\leq L \leq 510$. There exist weak oscillations of entropy around this linear line. The entropy at a large value $L=1000$ is plotted to show the validity of the linear fit. The slope is found to be close to $1$, indicating that the central charge is $2$.   
    }
		\label{ee}
	\end{figure} 

{\it DMRG numerics.} In this section,  we use DMRG from ITensor\cite{itensor} to simulate the lattice model in Eq.~\ref{1}. We fix model parameters to $t_0=t_2=1$ and $t_1=2$, and the filling is half. The truncation error cutoff is taken to be $10^{-10}$. We keep running sweeps until $E_i-E_{i-1}<10^{-11}$ and $S_i-S_{i-1}<10^{-8}$. Here $E_i$ and $S_i$ are the energy and entropy of the ground state at $i$-th sweep. Below we consider different values of coupling constants in the lattice model to explore low-energy descriptions of interacting $z=2$ LCs. 

Firstly, we consider the situation with positive $u_0$. In the previous section, we show the linear dispersion of the particle-hole collective mode. Therefore, one shall expect a logarithmic entanglement entropy $S$ for $u_0>0$. Here we plot the behavior of $u_0=0.1$ in Fig.~\ref{ee}. One can see that the entropy at $u_0=0.1$ climbs quickly and deviates from behaviors of the free $z=2$ LC. At larger $L$, a clear CFT characteristic emerges. We show that the central charge of the theory is $c=2$, which is plotted in the inset of Fig.~\ref{ee}. We also find that the finite-size gap at $u_0=+0.1$ is fundamentally linear in $1/L$.

\begin{figure} 
		\centering
		\includegraphics[scale=0.5]{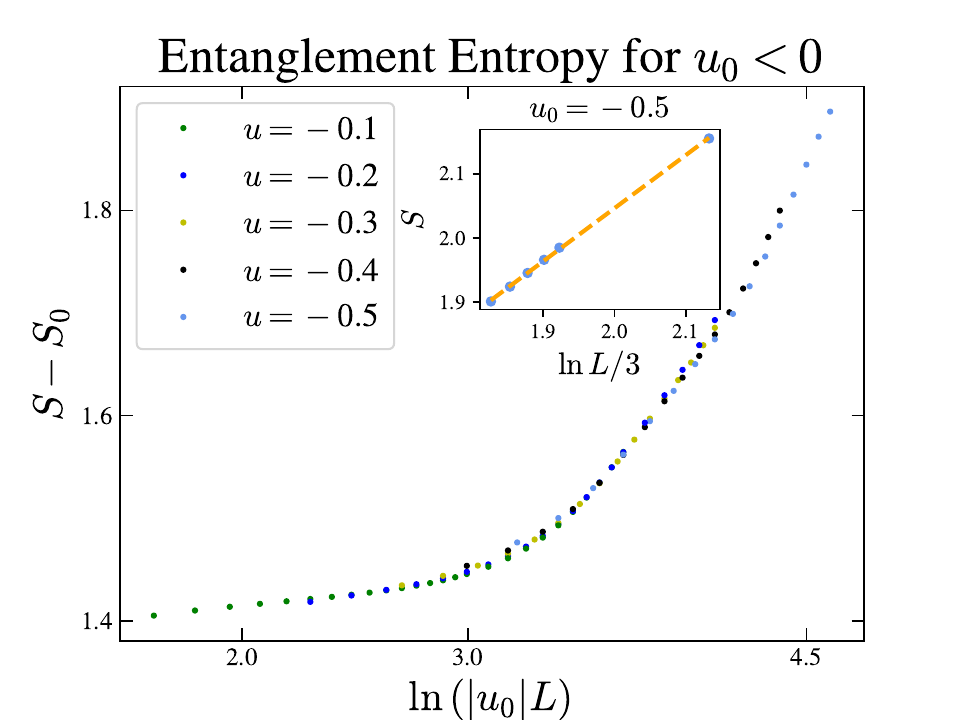} 
		\caption{ (Color Online)
  Data collapse of Entanglement entropy $S-S_0$ versus $u_0 L$ at $u_0=- 0.1,...,-0.5$. Here $S_0$ is some non-universal shift depending on $u_0$, numerically found $\sim u_0/8$.  The inset shows the entropy for $u_0=-0.5$ from $L=240$ to $320$ and its linear fit. A larger value $L=600$ is plotted to ensure the validity of linear fit. The slope is around $0.9(0)$.}
		\label{collapse_0}
	\end{figure}

Fig.~\ref{ee} also plots the entropy of $u_0=-0.1$. We observe that, up to the size of $L=200$, it only deviates from the behavior of the free case slightly. To study the negative $u_0$ more carefully, we plot $S$ of different negative $u_0=-0.1,...,-0.5$ and include the larger size of the system in Fig.~\ref{collapse_0}. One can observe a clear data collapse of $S$ v.s. $\ln u_0 L$. We use the $u_0=-0.5$ to see that the $S(L)$ converges to a linear function of $\log(L)$ and the slope is around $1/3$, indicating a CFT behavior again. Another observation is that $S$ converges to 
a linear function in $\log(L)$ when $-u_0 L\sim 120$. This is a huge scale, contrasted to $ u_0 L\sim 10\, (u_0>0)$ where the linear behavior has emerged. This observation indeed agrees well with RG flow in Eq.~\ref{9}. Since the generated $v$ is roughly proportional to $u_1$, the $u_1$ starting from positive $u_0$ increase much quicker than from negative $u_0$.

Another question pertains to the dynamical critical exponent $z$ of the new fixed point. We determine $z$ by computing the finite-size gap of the critical system and analyzing its scaling behavior for $1/L$, where $L$ denotes the system size. Here
 we consider
\begin{eqnarray}
\Delta   =E_1(L)-E_{0}(L) 
\end{eqnarray}
where $E_{1/0}$ is the energy of the first excitation/ground state. We see a clear data collapse between $L \Delta/f(u_0)$ and $Lu_0$. Here $f(u_0)$ is a non-universal function of $u_0$. Numerically we expand it as $u_0-\lambda u_0^2$ and $\lambda \simeq 1/25$, since $u_0<1$. We observe that $L \Delta/f(u_0)$ converges to a constant $\sim 0.3$ when $Lu_0$ approaches $120$. This indicates that the finite size gap scales $\Delta \sim 0.3 f(u_0)/L$. Therefore, the dynamical critical exponent at negative $u_0$ becomes $z=1$. This observation fits our RG flow well.

\begin{figure} 
		\centering
		\includegraphics[scale=0.5]{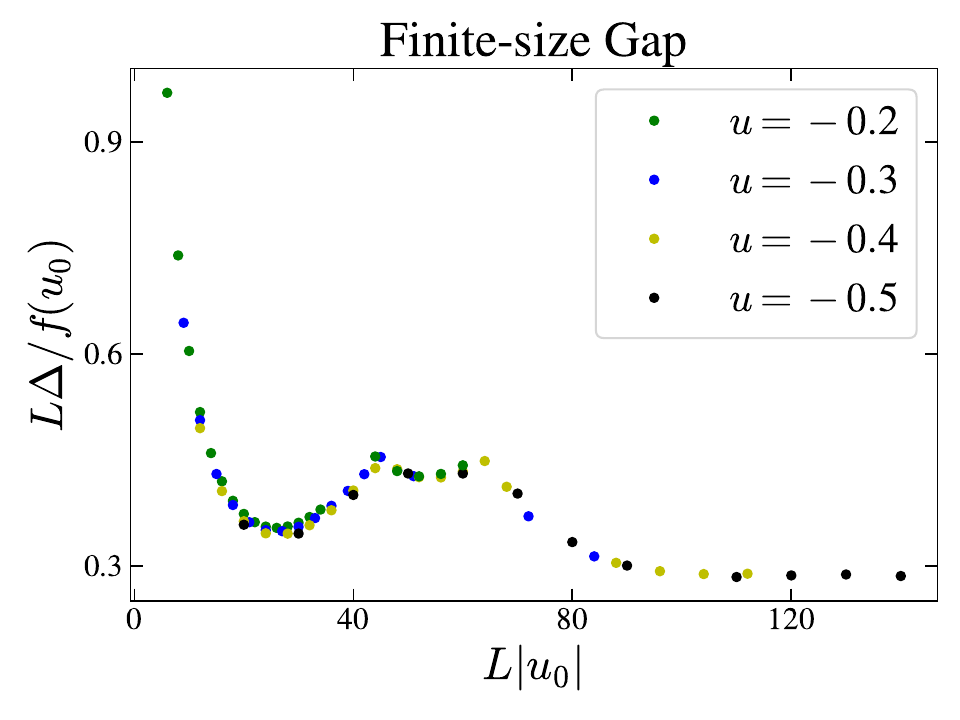} 
		\caption{ (Color Online)
  Data collapse for finite-size gap at $u_0<0$: $L\Delta /f(u_0)$ v.s. $Lu_0$. Here $f(u_0)$ is a function of $u_0$. Since $u_0$ is small here, we write $f(u_0)$ up to the second order of $u_0$ as $|u_0|-\lambda u_0^2$. Here $\lambda$ is the numerical coefficient around $1/25$. When $Lu_0>100$, a clear plateau emerges, indicating the finite-size gap scales as $\sim 1/L$. 
    }
		\label{energy}
	\end{figure}

{\it Conclusion and Remarks.}   In this work, we study the $z=2$ field theory in the one-dimensional interacting fermions with a Dirac Sea both analytically and numerically. Perturbative calculations indicate that the lifetime of a single particle vanishes. For the positive coupling constant, we obtain consistent results from RPA calculation and numerical simulations: the low-energy theory is described by CFT with $c=2$ and stable low-energy excitations are gapless bosons. Numerical simulations confirm this prediction.

For the negative coupling constant, the RPA result shows that there are no sharply well-defined collective modes. We show that there may exist a non-trivial RG fixed point characterized by two non-zero coupling constants. But if one only turns on $u_0$ at the UV scale, the velocity term is generated in the RG flow and dramatically changes the scaling of the quadratic Lagrangian. 

Our next step is to numerically scan the phase diagram versus hopping parameters/the on-site interaction to check whether there exists a non-trivial RG fixed point and further study its properties.


{\it Acknowledgment.}   K.W. thanks helpful suggestions on RG from the anonymous referee. K.W. thanks helpful discussions of DRMG with Jin Zhang. K.W. also thanks valuable discussions with D. T.~Son, L. Delacretaz,  X. C. Wu, K. Levin, T. Sedrakyan,  Y. H. Zhang, W. Witczak-Krempa, A. Klein,   Z. Q. Wang, U. Mehta, and A. Hui.    The author  is supported, in part, by the Kadanoff Center for Theoretical Physics.

\bibliography{sample}

 \pagebreak
 
 \widetext
 \begin{center}
 	\textbf{\large Supplementary material for Quantum $z=2$ Lifshitz criticality in one-dimensional interacting fermions }
 \end{center}
 \setcounter{equation}{0}
 \setcounter{figure}{0}
 \setcounter{table}{0}
 \setcounter{page}{1}
 \makeatletter
 \renewcommand{\theequation}{S\arabic{equation}}
 \renewcommand{\thefigure}{S\arabic{figure}}
 \renewcommand{\bibnumfmt}[1]{[S#1]}
 \renewcommand{\citenumfont}[1]{S#1}

\section{Perturbative calculations} \label{integral}

\subsection{Derivation of Polarization Operator }
We start the definition of the polarization operator,
\begin{eqnarray}
\Pi_{s,s'}(q,\omega)&=&-i  \int \frac{dk}{2\pi}\frac{d\omega'}{2\pi}G^s_0(k,\omega') G^{s'}_0(k-q,\omega'-\omega).
\end{eqnarray}
Insert the expression of the propagators,
\begin{eqnarray}
\Pi_{s,s'}(q,\omega)&=&-i   \int \frac{dk}{2\pi}\frac{d\omega'}{2\pi} \frac{1}{\omega'-s\epsilon(k)+i\delta \text{sgn}(\omega')} \frac{1}{\omega'-\omega-s'\epsilon(k-q)+i\delta \text{sgn}(\omega'-\omega)}.
\end{eqnarray}
Here $\epsilon(k)=k^2$. Note that in the integral, the effect of $\text{sgn}(\omega)$ is same as the $s$. Therefore we define 
\begin{eqnarray}
f_{s,s'}(k,q)=\int_{-\infty}^{\infty} \frac{d\omega'}{2\pi} \frac{1}{\omega'-s\epsilon(k)+is\delta } \frac{1}{\omega'-\omega-s'\epsilon(k-q)+i s' \delta  }
\end{eqnarray}
Now it is purely about applying the residue theorem in the complex analysis. That gives
\begin{eqnarray}
&&\text{if } s=s',\quad f_{s,s'}(k,q)=0\\
&&\text{if } s=- s', \quad f_{s,s'}(k,q)= is \frac{1}{-\omega+s\epsilon(k)+s\epsilon(k-q)-i s\delta}=-i  \frac{1}{s\omega- \epsilon(k)- \epsilon(k-q)+i  \delta}\nonumber
\end{eqnarray}
Thus one reaches the integral expression,
\begin{eqnarray}
\Pi_{s,s'}(q,\omega)=-(\sigma^{s,s'}_x)\int \frac{dk}{2\pi}  \frac{1}{s\omega- \epsilon(k)- \epsilon(k-q)+i  \delta}.
\end{eqnarray}
One may use $k^2+(k-q)^2=2(k-q/2)^2+q^2/2$ and redefine the integral variable $k'=k-q/2$,
\begin{eqnarray}
\label{17}
\Pi_{s,-s}(q,\omega)
&=&- \int_{-\infty}^{+\infty} \frac{dk}{ 2\pi} \frac{1}{ s\omega-2(k-q/2)^2- q^2/2+i \delta}=\int_{-\infty}^{+\infty} \frac{dk'}{ 4\pi} \frac{1}{k'^2+q^2/4-s \omega/2-i \delta } \nonumber.
\end{eqnarray}
Below we give an analysis based on the sign of  $q^2/4-s\omega/2 $.
\begin{itemize}
    \item If $q^2/4-s\omega/2u> 0$, $\delta$ is not important and can be taken as zero. One may need the equation below,
\begin{eqnarray}
\int_{-\infty}^{+\infty} \frac{dz}{2\pi} \frac{1}{z^2+ a^2}=\int_{-\infty}^{+\infty} \frac{dz}{2\pi} \frac{1}{(z+ia)(z-ia)}
\end{eqnarray}
Deforming the contour to the upper arc, one finds
\begin{eqnarray}
\Pi_{s,-s}(q,\omega)
&=& \frac{1}{ 2 \sqrt{ -2s\omega+ q^2} },\quad  \text{ if } q^2/4+s\omega/2u> 0
\end{eqnarray}

  \item If $q^2/4-s\omega/2u< 0$, the PO may be rewritten as
  \begin{eqnarray}
  \Pi_{s,-s}(q,\omega)
&=&\int_{-\infty}^{\infty} \frac{dk}{ 4\pi} \frac{1}{k^2-a^2-i \delta } = -\frac{i}{4 } \frac{1}{ \sqrt{|-s\omega/2 + q^2/4|}  }
  \end{eqnarray}
 where $a^2=|q^2/4-s \omega/2|$
\end{itemize}
Therefore, one can reach a simple expression,
\begin{eqnarray}
\Pi_{s,s'}(q,\omega)= \frac{1}{4} \frac{|s-s'|}{\sqrt{ - 2s\omega +  q^2 }}.
\end{eqnarray}
This is Eq.~4 in the main-text.
\subsection{One-loop Self energy: imaginary part}
Correction to the green function may be obtained from inserting the self energy,
\begin{eqnarray}
\delta G_+(k,\omega)= G_+(k,\omega) \Sigma_+(k,\omega) G_+(k,\omega).
\end{eqnarray}
At the one loop level, the self energy is approximated by
\begin{eqnarray}
\Sigma_+(k,\omega)\simeq -i u_0^2 \int \frac{d\nu}{2\pi }\int \frac{dq }{ 2\pi }    G_{-}(k-q,\omega-\nu)    \Pi_{+,-}(q,\nu)
\end{eqnarray}
Note that the other vertex-like diagram (usually called Fock diagram) in the second order perturbation does not contribute since the coupling constant does not contain the scattering process $++\rightarrow --$. Note that in the Fermi liquid theory, the Fock diagram contributes since it involves the backscattering events between two {\it different} points in the Fermi surface. However, we only have a single Fermi point here.
Now we trace the imaginary part of $\Sigma_+$. 
\begin{eqnarray}
\text{Im} \Sigma_+(k,\omega)\equiv F_1+F_2;
\end{eqnarray}
where $F_{1/2}$ involves the product of two imaginary/real parts of $G$ and $\Pi$,
\begin{eqnarray}
F_1&=&\int \frac{d\nu}{2\pi }\int \frac{dq }{ 2\pi }   \text{Im} G_{-}(k-q,\omega-\nu)      \text{Im}\Pi_{+,-}(q,\nu),\\
F_2&=& -\int \frac{d\nu}{2\pi }\int \frac{dq }{ 2\pi }   \text{Re} G_{-}(k-q,\omega-\nu)      \text{Re}\Pi_{+,-}(q,\nu).
\end{eqnarray}
Below we evaluate two integrals of $F_1$ and $F_2$.

{\bf Evaluation of  $F_1$ .} Note that $\text{Im}G_{-}(k-q,\omega-\nu)=  \pi\delta(\omega-\nu+ \epsilon_{k-q})$ and 
\begin{eqnarray}
\text{Im} \Pi_{+-}(q,\nu)=-\frac{1}{2\sqrt{2} \sqrt{  \nu - q^2/2 }}\Theta(  \nu - q^2/2)
\end{eqnarray}
Then one may obtain
\begin{eqnarray}
F_1= -\frac{  u_0^2  }{2\sqrt{2} }  \int \frac{dq }{ 2\pi }   \int \frac{d\nu}{2\pi }
\pi\delta(\omega-\nu+ \epsilon_{k-q}) \frac{1}{\sqrt{    \nu - q^2/2 }}\Theta( \nu - q^2/2). 
\end{eqnarray}
Integration over $\nu$ gives,
\begin{eqnarray}
F_1= -\frac{  u_0^2  }{4\sqrt{2}  }  \int \frac{dq }{ 2\pi }   
 \frac{1}{\sqrt{  \omega +  (k-q)^2- q^2/2 }}\Theta( \omega +  (k-q)^2- q^2/2) \nonumber
\end{eqnarray}
Now we see one special kind of term appearing above, $(k-q)^2-\frac{q^2}{2}$. We will see this term many times when we calculate the self-energy involved with one-loop diagrams. This special has a clear physical origin: poles of the polarization operator is {\it not} same as ones of propagators. The former one locates the minimum of particle-hole continuum while the latter one is the free fermion excitation.

The difference of two poles, $ (k-q)^2-\frac{q^2}{2}$, can be reformulated by $\frac{1}{2}(q-2k )^2-k^2$. Redefining the integration variable, one finds
     \begin{eqnarray}
      J_+=   -\frac{u_0^2}{4\sqrt{2} }  \int_{-\infty}^{\infty}  \frac{dq }{ 2\pi }  
 \frac{1 }{\sqrt{  q^2/2+  \omega    -k^2 }}\Theta(q^2/2+  \omega    -k^2).
     \end{eqnarray}
  There emerges one important quantity, $Q^2 \equiv  |\omega-k^2|$.   Now we consider two different scenarios based on the sign of $\omega -k^2$.
     \begin{itemize}

      \item {$    \omega - k^2 \geq 0 $}.  The integral reads  
     \begin{eqnarray}
     J_+=-\frac{u_0^2}{2  }  \int_{0}^{\infty} \frac{dq }{ 2\pi }  
 \frac{ 1}{\sqrt{    q^2 +Q^2 }} =-\frac{u_0^2}{4\pi }\ln  \frac{\Lambda+\sqrt{\Lambda^2+Q^2}}{Q}  ,\quad Q\geq 0
     \end{eqnarray}
     where $\Lambda$ is UV cutoff. Note that there is no Fermi-energy as the upper bound of the momentum. We only need $\Lambda<\pi/a$ where $a$ is the lattice spacing. We may claim the result below
      \begin{eqnarray}
      J_+\simeq -\frac{u_0^2}{4\pi }\ln   
      \frac{ \Lambda}{\sqrt{\omega -k^2} }.
        \end{eqnarray}
 One may observe a {  blow-up} at the on-shell condition $\omega= k^2$. It is traced back to a infrared divergence in the integral. Even at off-shell case, $J_+$ is still logarithmically large .
    
     \item {$ \omega -k^2 \leq 0 $}. The integral reads
     \begin{eqnarray}
J_+= -\frac{u_0^2}{2  } \int_{Q}^{\infty} \frac{dq }{ 2\pi }  
  \frac{1}{ \sqrt{   q^2 -Q^2 }} =-\frac{u_0^2}{4\pi } \ln\left(\frac{\Lambda}{\sqrt{|\omega-k^2|}} \right)
\end{eqnarray}
\end{itemize}
As a summary of $J_+$, we find the expression valid for arbitrary $\omega -k^2$,
   \begin{eqnarray}
   \boxed{
   J_+\simeq -\frac{u_0^2}{4\pi}\ln   
      \frac{ \Lambda}{\sqrt{|\omega-k^2|}}.
      }
 \end{eqnarray}
 
{\bf Evaluation of $F_2$.}
In the previous calculation, we only focus on the imaginary part. 
The real part of polarization becomes
\begin{eqnarray}
 \text{Re} \Pi_{+-}(q,\nu)= \frac{1}{2\sqrt{2}\sqrt{   q^2/2 -   \nu  }} \Theta(  q^2/2 -   \nu)
\end{eqnarray}
The real part of the propagator carries the  principal part,
\begin{eqnarray}
\text{Re}G_{-}(k-q,\omega-\nu)=P \frac{1}{\omega-\nu+  (k-q)^2 }.
\end{eqnarray}
Defining a new variable $x=q^2/2-\nu $, one may rewrite the denominator in the propagator as
\begin{eqnarray}
\omega-\nu+  (k-q)^2=x+X(q),\quad X(q)\equiv \omega+  k^2-2kq+q^2/2=\frac{1}{2}(q-2k)^2+\omega-k^2.
\end{eqnarray}
With a proper redefinition of $q$, one obtain the expression of $F_2$
 \begin{eqnarray}
F_2\simeq   -u_0^2\int \frac{dx}{2\pi }\int \frac{dq }{ 2\pi }   P \frac{1}{x+\frac{1}{2}q^2+\omega-k^2} \frac{1}{2\sqrt{2} \sqrt{ x}} \Theta(x ).
\end{eqnarray}
One need to consider two different scenarios.
\begin{itemize}
    \item  {$    \omega - k^2 \geq 0 $}.    
    \begin{eqnarray}
F_2\simeq   -\frac{u_0^2}{\sqrt{2}} \int \frac{dx}{2\pi }\int_{-\infty}^{\infty}  \frac{dq }{ 2\pi }    \frac{1}{  q^2+2x+2Q^2} \frac{1}{ \sqrt{ x}} \Theta(x )=- \frac{u_0^2}{4}\int^{+\infty}_0 \frac{dx}{2\pi }   \frac{1}{ \sqrt{ x+ Q^2} } \frac{1}{\sqrt{ x}} 
\end{eqnarray}
We use the UV cut-off to regulate the integral,
 \begin{eqnarray}
 \int^{+\Lambda^2}_0 \frac{dx}{2\pi }   \frac{1}{ \sqrt{ x+ Q^2} } \frac{1}{\sqrt{ x}}\simeq \frac{1}{\pi} \ln \left(
 \frac{2\Lambda}{Q}
 \right)
 \end{eqnarray}
 Therefore, $F_2$ is given by
 \begin{eqnarray}
 \label{S29}
 F_2=- \frac{u_0^2}{4\pi }  \ln \left(
 \frac{2\Lambda}{Q}
 \right)
  \end{eqnarray}

   \item  {$    \omega - k^2 < 0 $}.   \begin{eqnarray}
F_2\simeq   -2u_0^2\int \frac{dx}{2\pi }\int \frac{dq }{ 2\pi }    P \frac{1}{ q^2+2x- 2Q^2} \frac{1}{2\sqrt{2} \sqrt{ x}} \Theta(x )
\simeq - \frac{u_0^2}{4}\int^{+\infty}_{Q^2}\frac{dx}{2\pi }   \frac{1}{ \sqrt{ x- Q^2} } \frac{1}{\sqrt{ x}} 
\end{eqnarray}   
In the last "equality", we approximation $F_2$ by ignoring the contribution from $x<Q^2$. Not hard to see that the integral gives the same integral with Eq.~\ref{S29}.

\end{itemize}
As a conclusion, we find that the imaginary of self-energy contains a logarithmic divergence,
\begin{eqnarray}
\text{Im} \Sigma_+(k,\omega)=-  \frac{u^2_0}{2\pi}  \ln \frac{\Lambda}{\sqrt{|\omega-k^2|}}
\end{eqnarray}
which is Eq.~5 in the maintext.
\section{RPA Self-energy}
Consider the RPA self energy for particle-like excitations,
\begin{eqnarray}
\Sigma_{\text{RPA},+}(k,\omega)\simeq -i u_0 \int \frac{d\nu}{2\pi }\int \frac{dq }{ 2\pi }    G_{-}(k-q,\omega-\nu)    \frac{1}{1-u_0 \Pi_{+,-}(q,\nu) }
\end{eqnarray}
\subsection{Real Part of $\Sigma_{\text{RPA} }$}
Now we would like to trace its real part.
\begin{eqnarray}
\text{Re} \Sigma_{\text{RPA},+}(k,\omega)=u_0 \text{Im}\int \frac{d\nu}{2\pi }\int \frac{dq }{ 2\pi }    G_{-}(k-q,\omega-\nu)    \frac{1}{1-u_0 \Pi_{+,-}(q,\nu) }
\end{eqnarray}
Again, one may separate it into two terms,
\begin{eqnarray}
R_1=u_0
\int \frac{d\nu}{2\pi }\int \frac{dq }{ 2\pi }    \text{Im}G_{-}(k-q,\omega-\nu)   \text{Re} \frac{1}{1-u_0 \Pi_{+,-}(q,\nu) }, \label{ss37}
\\ R_2=u_0
\int \frac{d\nu}{2\pi }\int \frac{dq }{ 2\pi }    \text{Re}G_{-}(k-q,\omega-\nu)   \text{Im} \frac{1}{1-u_0 \Pi_{+,-}(q,\nu) }.
\end{eqnarray}
 
 {\bf Contribution from $R_1$. } Recall that $\text{Im}G_{-}(k-q,\omega-\nu)$ reads $\pi \delta(\omega-\nu+ \epsilon_{k-q})$. Then one may write Eq.~\ref{ss37} as
\begin{eqnarray}
R_1=\pi u_0
\int \frac{d\nu}{2\pi }\int \frac{dq }{ 2\pi } \delta(\omega-\nu+ \epsilon_{k-q})\left( 
 \frac{ \Theta(-   \nu +  q^2/2 )}{1-    \frac{ u_0}{2\sqrt{ - 2 \nu +  q^2 }}  }  +    \text{Re}\frac{\Theta(    \nu -  q^2/2 )}{1+   i \frac{ u_0}{2\sqrt{   2 \nu -  q^2 }}   } 
\right)
\end{eqnarray}  
The delta function set $\nu=\omega+\epsilon_{k-q}$. We see the combination $(k-q)^2-\frac{q^2}{2}$ again. As usual we use  $2(k-q)^2-q^2= (q-2k )^2-2 k^2$ and replace the integral variable $q$ by $q'=q-2k$. This gives
\begin{eqnarray}
R_1&=&\frac{u_0}{2}
 \int \frac{dq' }{ 2\pi } \left( 
 \frac{ \Theta( - 2 \omega -  q'^2+2k^2 )}{1-    \frac{ u_0}{2\sqrt{ - 2 \omega -  q'^2+2k^2 }}  }  +  \text{Re} \frac{\Theta(   2 \omega + q'^2-2k^2 )}{1+   i \frac{ u_0}{2\sqrt{    2 \omega + q'^2-2k^2 }}   } 
\right)
\end{eqnarray} 
Again, one has to study two different situations based on the sign of $ 2k^2-2\omega$.   
\begin{itemize}
    \item {Case 1: $\omega<k^2$.} In this case,   
\begin{eqnarray}
R_1=  u_0
 \int^{\sqrt{2k^2-2\omega}}_{0} \frac{dq' }{ 2\pi } 
 \frac{1}{1-    \frac{ u_0}{2\sqrt{ - 2 \omega -  q'^2+2k^2 }}   }  +   u_0
 \int_{ \sqrt{2k^2-2\omega}}^{+\Lambda} \frac{dq' }{ 2\pi }   \frac{1}{1+ \frac{1}{4}\frac{ u^2_0 }{q'^2+     2 \omega   -2k^2  }   } 
\end{eqnarray} 
Let us express the term above in the momentum scale $Q=\sqrt{2k^2-2\omega}$,
\begin{eqnarray}
R_1= u_0 \left(
 \int^{Q}_{0} \frac{dq' }{ 2\pi } 
 \frac{1}{1-    \frac{ u_0}{2\sqrt{ Q^2 -  q'^2  }}   }  +  
 \int_{ Q}^{+\Lambda} \frac{dq' }{ 2\pi }    \frac{1}{1+ \frac{1}{4}\frac{ u^2_0 }{q'^2-Q^2  }   } \right) 
\end{eqnarray} 
{\bf Case 1a: $u_0^2/4-Q^2>0$.} 
One may use the indefinite integral below,
\begin{eqnarray}
\int dx \frac{1}{1-b/\sqrt{a^2-x^2}}=-\frac{b^2}{\sqrt{b^2-a^2}} \tan^{-1}\left(\frac{bx}{\sqrt{b^2-a^2} \sqrt{a^2-x^2}  } \right)-\frac{b^2}{\sqrt{b^2-a^2}}  \tan^{-1} \left( \frac{x}{\sqrt{b^2-a^2}}\right)+b\tan^{-1}\frac{x}{\sqrt{a^2-x^2}}+x \nonumber
\end{eqnarray} 
to evaluate the first integral.
Setting $b=u_0/2$ and $a=Q$, one reach
\begin{eqnarray}
\label{39}
 \int^{Q}_{0} \frac{dq' }{ 2\pi } 
 \frac{1}{1-    \frac{ u_0}{2\sqrt{ Q^2 -  q'^2  }}   }=-\frac{u_0^2}{8\pi}\frac{\pi/2 \text{sgn}(u_0)+\arctan (Q/\sqrt{u^2_0/4-Q^2}) }{\sqrt{u^2_0/4-Q^2}}+ \frac{u_0}{4\pi}\frac{\pi}{2}+[u_0\text{-independent}]
\end{eqnarray} 
The second integral reads
\begin{eqnarray}
 \int_{ Q}^{+\Lambda} \frac{dq' }{ 2\pi }    \frac{1}{1+ \frac{1}{4}\frac{ u^2_0 }{q'^2-Q^2  }   } = -    \int_{ Q}^{+\Lambda} \frac{dq' }{ 2\pi }   \frac{u_0^2/4}{q'^2-Q^2+u_0^2/4}+[u_0\text{-independent}]. 
\end{eqnarray} 
Here we usually ignore the $u_0\text{-independent}$ contribution in the integral. Reason is following: they give corrections to the self-energy linear in $u_0$ which is fundamentally tap-dole diagram contributions. These contributions also do not contain any $k,\omega$-dependence. One may use the formula below
\begin{eqnarray}
\int dx \frac{1}{x^2+a^2}= \frac{ 1}{a} \tan^{-1}\left(
\frac{x}{a}
\right)
\end{eqnarray} 
 to estimate the second integral.
Notice that the denominator is always positive due to $u_0^2$. Setting $a^2=u_0^2/4-Q^2$, the integral gives
\begin{eqnarray}
- \int_{ Q}^{+\Lambda} \frac{dq' }{ 2\pi }   \frac{u^2_0/4}{q'^2-Q^2+u_0^2/4}\simeq -\frac{u_0^2 }{16\sqrt{u_0^2/4-Q^2}}+ \frac{u_0^2 }{8\pi\sqrt{u_0^2/4-Q^2}} \arctan \frac{Q}{\sqrt{u_0^2/4-Q^2}}
\end{eqnarray} 
When $u_0>0$, one would see that a divergence at $Q^2=u_0^2/4$ in Eq.~\ref{39}. This is understood as the particle-hole collective modes.  Sum two integrals, one reaches 
\begin{eqnarray}
R_1=    \left(\frac{u^2_0  }{8}-(\text{sgn}(u_0) +1)\frac{u_0^3 }{16\sqrt{u_0^2/4-Q^2}} \right) \Theta(u_0^2/8-|k^2-\omega|) \Theta(k^2-\omega).
\end{eqnarray} 
If $u_0>0$, the on-shell condition gives $\omega=k^2+ \frac{u^2_0  }{8}-\frac{u_0^3 }{8\sqrt{u_0^2/4-Q^2}}$, causing emergence of Fermi-velocity. 

{\bf Case 1b: $u_0^2/4-Q^2<0$. } Consider the first integral. One may use the formula 
\begin{eqnarray}
\arctan ix=\frac{i}{2}\log \left(\frac{1+x}{1-x}\right)
\end{eqnarray}
One may need to evaluate
\begin{eqnarray}
&&-\frac{b^2}{\sqrt{b^2-a^2}} \Big[ \tan^{-1}\left(-i  u_0 \infty   \right) +   \tan^{-1} \left(-i \frac{a}{\sqrt{|b^2-a^2|}}\right)\Big]+b \frac{\pi}{2}
\\
&=&-\frac{b^2}{\sqrt{b^2-a^2}} \Big[ \frac{i}{2}\log\left(\frac{1-u_0 \infty}{1+u_0 \infty}\right) +    \frac{i}{2}\log \left(\frac{\sqrt{b^2-a^2}-a}{\sqrt{b^2-a^2}+a}\right)   \Big]+b \frac{\pi}{2} 
-\frac{b^2}{\sqrt{|b^2-a^2|}} \Big[     \frac{1}{2}\log \left(\frac{a-\sqrt{b^2-a^2} }{\sqrt{b^2-a^2}+a}\right)   \Big]+b \frac{\pi}{2} \nonumber
\end{eqnarray}
Then the first integral in $R_1$ gives
\begin{eqnarray}
\int^{Q}_{0} \frac{dq' }{ 2\pi } 
 \frac{1}{1-    \frac{ u_0}{2\sqrt{ Q^2 -  q'^2  }}   }= 
  \frac{u_0^2}{ 16\pi \sqrt{|u_0^2/4-Q^2|}}  \log \left(  
 \frac{Q+\sqrt{|u_0^2/4-Q^2|}}{Q-\sqrt{|u_0^2/4-Q^2|}}
 \right) +\frac{  u_0}{8 }
\end{eqnarray} 
The second integral in $R_1$ gives
\begin{eqnarray}
-   \int_{ Q}^{+\Lambda} \frac{dq' }{ 2\pi }   \frac{u_0^2/4}{q'^2-Q^2+u_0^2/4}\simeq  -\frac{u_0^2}{16\pi \sqrt{|u_0^2/4-Q^2|}} \ln \left(  
 \frac{Q+\sqrt{|u_0^2/4-Q^2|}}{Q-\sqrt{|u_0^2/4-Q^2|}}
 \right).
\end{eqnarray} 
In this case, one may write down the epxression of $R_1$ as
\begin{eqnarray}
R_1=\frac{u^2_0   }{8} \Theta( k^2-\omega -u_0^2/8), \quad  \text{if } u_0^2/4<2|k^2-\omega|
\end{eqnarray}

\item {Case 2: $\omega>k^2$.}  In this case expression is much simpler,
\begin{eqnarray}
R_1=  u_0  
 \int_{ 0}^{+\Lambda} \frac{dq' }{ 2\pi }    \frac{1}{1+ \frac{1}{4}\frac{ u^2_0 }{q'^2+Q^2  }   }
\end{eqnarray} 
Here we define $Q=\sqrt{2\omega-2k^2}$. This integral can be easily integrated out,
\begin{eqnarray}
\int_{ 0}^{+\Lambda} \frac{dq' }{ 2\pi }  \frac{1}{1+ \frac{1}{4}\frac{ u^2_0 }{q'^2+Q^2  }   }&=&\int_{ 0}^{+\Lambda} \frac{dq' }{ 2\pi } \frac{q'^2+Q^2 }{q'^2+Q^2 + \frac{1}{4} u_0^2   }=\int_{ 0}^{+\Lambda} \frac{dq' }{ 2\pi } \left(1-\frac{u_0^2/4}{q'^2+Q^2 + \frac{1}{4} u_0^2   } \right)\\
&=&\Lambda/2\pi-\frac{u_0^2}{8 }\int_{ \mathbb{R}} \frac{dq' }{ 2\pi }  \frac{1}{q'^2+Q^2 + \frac{1}{4} u_0^2   }=\Lambda/2\pi-\frac{u_0^2}{8 }    \frac{1}{2\sqrt{Q^2+u_0^2/4}}
\end{eqnarray} 
Again we ignore the constant term in the integral. One may obtain the contribution to the self-energy 
\begin{eqnarray}
R_1=-  \frac{ u_0^3}{16}  \frac{1}{ \sqrt{2\omega-2k^2+u_0^2/4}} \Theta(\omega-k^2).
\end{eqnarray}
If we set $\omega=k^2$, then it reduces to be $-\text{sgn}(u_0)\pi u_0^2/4$. When $u_0$ is negative, result is consistent with the one from the calculation in the case $\omega<k^2$.  
Therefore we can unify the expression of the $R_1$ to be
\begin{eqnarray}
\frac{u_0^2}{8}\Theta(k^2-\omega) -(\text{sgn}(u_0) +1)\frac{u_0^3 }{16\sqrt{u_0^2/4-2k^2+2\omega}}  \Theta(u_0^2/8-|k^2-\omega|) \Theta(k^2-\omega)-  \frac{ u_0^3}{16}  \frac{1}{ \sqrt{2\omega-2k^2+u_0^2/4}} \Theta(\omega-k^2) \nonumber
\end{eqnarray}


\end{itemize}
 
{\bf Contribution from $R_2$. } 
From its definition, one may write down  the expression of $R_2$,
\begin{eqnarray}
R_2&=&  u_0\int \frac{d\nu}{2\pi }\int \frac{dq }{ 2\pi }  
P \frac{1}{\omega-\nu+   (k-q)^2 }
   \text{Im}\frac{\Theta(    \nu -  q^2/2 )}{1+   i \frac{ u_0}{2\sqrt{   2 \nu -  q^2 }}   } 
  \end{eqnarray} 
Performing Im, one reaches
\begin{eqnarray}
R_2 
   &=& -u_0
\int \frac{d\nu}{2\pi }\int \frac{dq }{ 2\pi }  
P \frac{1}{\omega-\nu+   (k-q)^2 }
    \frac{\Theta(    \nu -  q^2/2 )}{1+    \frac{ u^2_0}{8 \nu -  4 q^2  }    } \frac{ u_0}{2\sqrt{   2 \nu -  q^2 }} 
\end{eqnarray}
Defining a new variable $x=-q^2/2+\nu $, one may rewrite the denominator in the propagator as
\begin{eqnarray}
\omega-\nu+  (k-q)^2=-x+X(q),\quad X(q)\equiv \omega+  k^2-2kq+q^2/2=\frac{1}{2}(q-2k)^2+\omega-k^2.
\end{eqnarray}
With a proper redefinition of $q$, one obtain the expression of $R_2$
 \begin{eqnarray}
R_2\simeq   -u_0 \int \frac{dx}{2\pi }\int \frac{dq' }{ 2\pi }   P \frac{1}{-x+q'^2+\omega-k^2} \frac{\Theta(    x )}{1+    \frac{ u^2_0}{8 x  }    } \frac{ u_0}{2\sqrt{  x}} .
\end{eqnarray}
We consider the integration over $q$ firstly. 
\begin{itemize}
    \item If $\omega-k^2<0$, then the integral below vanishes,
    \begin{eqnarray}
 \int_{q\in \mathbb{R}} \frac{dq}{ 2\pi }  
P \frac{1}{q^2  -a^2  }
    =\int_{q\in \mathbb{R}}  \frac{dq}{ 2\pi } \frac{1}{2a} \left(P \frac{1}{q-a}-P \frac{1}{q+a}\right)=
  0.
\end{eqnarray}
Here $a$ is any real number. Therefore $R_2=0$ in this case.
 \item If $\omega-k^2>0$, one has to handle the integral
 \begin{eqnarray}
\int \frac{dq  }{ 2\pi }  P \frac{1}{q^2+ a^2}=\frac{1}{2 a}
 \end{eqnarray}
 Notice that no poles exist. Therefore one can ignore the principal part here. Therefore
 $R_2$ reduces to be
 \begin{eqnarray}
R_2\simeq   -\frac{u_0^2}{4} \int_0^{b} \frac{dx}{2\pi } \frac{1 }{1+    \frac{ u^2_0}{8 x  }    } \frac{1}{\sqrt{  x}}  \frac{1}{ \sqrt{b-x}}.
\end{eqnarray}
Here we denote $b=\omega-k^2$. One may use the formula below,
\begin{eqnarray}
\int dx \frac{1 }{1+    a/x    } \frac{1}{\sqrt{  x}}  \frac{1}{ \sqrt{b-x}}=2 \tan^{-1} \left( \frac{\sqrt{x}}{\sqrt{b-x}} \right)- \frac{2\sqrt{a}   }{\sqrt{a+b}} \tan^{-1}\left( \frac{(a+b)x}{a(b-x)}\right).
\end{eqnarray}
Setting $a=u_0^2/8$, one obtain the expression of $R_2$,
\begin{eqnarray}
R_2\simeq   -\frac{u_0^2}{8 } \left( 
1- \frac{ |u_0|/2}{\sqrt{2\omega-2k^2+u_0^2/4}}
\right)=\left(-\frac{u_0^2}{8}+ \frac{ u^3_0 \text{sgn}(u_0)}{16 }\frac{ 1}{\sqrt{2\omega-2k^2+u_0^2/4}}\right) \Theta(\omega-k^2)
\end{eqnarray}
\end{itemize}
Recall $R_1$ is given by
$$
\frac{u_0^2}{8}\Theta(k^2-\omega) -(\text{sgn}(u_0) +1)\frac{u_0^3 }{16\sqrt{u_0^2/4-2k^2+2\omega}}  \Theta(u_0^2/8-|k^2-\omega|) \Theta(k^2-\omega)-  \frac{ u_0^3}{16}  \frac{1}{ \sqrt{2\omega-2k^2+u_0^2/4}} \Theta(\omega-k^2)
$$
Summing $R_1+R_2$, we arivie the RPA self energy,
\begin{eqnarray}
\text{Re}\Sigma_{\text{RPA}}=-\frac{u_0^2}{8}\text{sgn}(\omega-k^2)  - \frac{u_0^3 }{16\sqrt{u_0^2/4+2\omega-2k^2}}  \Big(  1-\text{sgn}(\omega-k^2)\text{sgn}(u_0) \Big)
\end{eqnarray}

\subsection{Imaginary Part of $\Sigma_{\text{RPA} }$}
Since we have evaluated the one-loop level imaginary self-energy, here we only aim to check whether the functional form is affected in the RPA level. Consider
\begin{eqnarray}
\text{Im} \Sigma_{\text{RPA},+}(k,\omega)=-u_0 \text{Re}\int \frac{d\nu}{2\pi }\int \frac{dq }{ 2\pi }    G_{-}(k-q,\omega-\nu)    \frac{1}{1-u_0 \Pi_{+,-}(q,\nu) }.
\end{eqnarray}
Now we look at the piece which represents the physical process that a single-particle decays into particle-hole continuum,
\begin{eqnarray}
\label{s37}
I_1=u_0
\int \frac{d\nu}{2\pi }\int_{-\infty}^{\infty} \frac{dq }{ 2\pi }    \text{Im}G_{-}(k-q,\omega-\nu)   \text{Im} \frac{1}{1-u_0 \Pi_{+,-}(q,\nu) }
\end{eqnarray}
One repeats the procedure we have done in the one-loop calculation and finds
\begin{eqnarray}
I_1=\frac{u_0}{2} \int_{-\infty}^{\infty}  \frac{dq }{ 2\pi }   
\text{Im}\frac{1}{1+i    
 \frac{u_0 }{2\sqrt{2}\sqrt{  q^2/2+  \omega    -k^2 }}  }\Theta(q^2/2+  \omega    -k^2)
\end{eqnarray}
Expanding the denominator to the first order in $u_0$, one can see the one-loop expression $F_1$. For simplicity, we consider $\omega-k^2\geq 0$.  It gives
\begin{eqnarray}
I_1=-\frac{u_0}{2} \int_{-\infty}^{\infty}  \frac{dq }{ 2\pi }  \frac{u_0 }{2\sqrt{2}\sqrt{  q^2/2+  \omega    -k^2 }}   \times  
\frac{1}{1+    
 \frac{u^2_0 }{8  (q^2/2+  \omega    -k^2)  }  }=- \frac{u_0}{2}   \int_{0}^{\Lambda}  \frac{dq' }{ 2\pi }      
\frac{\sqrt{  q^{'2}+  \omega    -k^2 }}{ q^{'2}+  \omega    -k^2 +    
 \frac{u^2_0 }{8     }  }.
\end{eqnarray}
In the first equality, we take the imaginary part of the integrand. In the second equality, we define $q'=q/\sqrt{2}.$  One may check the integral table and use the indefinite integral below,
\begin{eqnarray}
\int dx \frac{\sqrt{x^2+a^2}}{x^2+a^2+b^2}=\log\left(\sqrt{a^2+x^2}+x \right)
-\frac{b}{\sqrt{a^2+b^2}}\tanh^{-1} \left(
\frac{bx}{\sqrt{a^2+b^2} \sqrt{a^2+x^2}}
\right).
\end{eqnarray}
Setting $a^2=\omega-k^2$ and $b^2=u_0^2/8$, one may reach
\begin{eqnarray}
  I_1\simeq - \frac{u^2_0 }{4 \pi }  \ln \left(\frac{\Lambda}{\sqrt{|\omega-k^2|}}\right)+\text{Converging Terms}.
\end{eqnarray}
Therefore, we confirm that the logarithmic divergence in the imaginary part of self remains valid in the RPA level.
    
  \section{Momentum-Shell Renormlaization group}
In the momentum-shell method, the RG is implemented by:  (1), integrating out the fast modes (2) obtaining the effective Lagrangian (3) Introduce the re-scaled parameters and field operators (which are used to keep Guassian action invariant). 

Now we have a bare theory defined by Lagrangian and action. Separate the field operator into slow and fast modes 
\begin{eqnarray} 
\psi(k)=\psi_>(k)+\psi_<(k),\quad >:  \Lambda/s<k<\Lambda ,\quad <:     0<k<\Lambda/s.
\end{eqnarray}
The operator $\psi_>(k)$ is defined by $\psi_>(k)=\psi(k)$ if $\Lambda/s<k<\Lambda$, otherwise zero. Similar definition for $\psi_<(k)$. Then the bare action/Lagrangian can be written as
\begin{eqnarray}
Z=\int D[\psi_>] D[\psi_<] e^{S(\psi_>,\psi_<)},\quad S(\psi_>,\psi_<)=S_0(\psi_>)+S_0(\psi_<)+S_{\text{int}}(\psi_>,\psi_<)
\end{eqnarray}
Effective action of $\psi_<$ can be obtained by integrating out the fast modes,
\begin{eqnarray}
Z=\int D[\psi_<]e^{S_0(\psi_<)} \int D[\psi_>]   e^{S_0(\psi_>)} e^{S_{\text{int}}(\psi_>,\psi_<)}=Z_> \int D[\psi_<] e^{S_0(\psi_<)} \langle e^{S_{\text{int}}(\psi_>,\psi_<)} \rangle_>
\end{eqnarray}
This equation determines the effective Lagrangian of $\psi_<$.
The average of interacting exponential is performed for the Gaussian action of $\psi_>$. The technical part is about evaluating the perturbation of $S_{\text{int}}$, which follows from 
\begin{eqnarray}
\log \langle e^{S_{\text{int}}(\psi_>,\psi_<)} \rangle_> \simeq \langle S_{\text{int}} \rangle_>+ \frac{\langle  S_{\text{int}}^2 \rangle_>-\langle S_{\text{int}} \rangle^2_>}{2}+  O(S^3_{\text{int}} )
\end{eqnarray}
The result is generally valid from cumulant expansion. Here I have hided the dependence of $S$ on the field operators.  Then the effective Lagrangian/action can be obtained perturbatively,
\begin{eqnarray}
S_{\text{eff}}(\psi_<)\simeq S_0(\psi_<)+\langle S_{\text{int}} \rangle_>+ \frac{\langle  S_{\text{int}}^2 \rangle_>-\langle S_{\text{int}} \rangle^2_>}{2}+...
\end{eqnarray}
Here I only preserve up to the second order. One can perform higher order perturbation, which is fundamentally equivalent to plotting connected Feynman diagrams. Although the first/second orders are written separately, they may contribute to the same coupling constant. Similar behaviors may happen for higher perturbations. One shall be careful about arguing that high-order perturbations do not change results.

To study the RG flow of coupling constant in $S_{\text{eff}}(\psi_<)$, one has to introduce the new momentum/frequency and field operators,
\begin{eqnarray}
\label{new_v}
k'=s k,\quad  \omega'=s^z \omega, \quad \psi'(k')=s^{-(2z+1)/2}\psi_<(k)
\end{eqnarray}
Here $k/\omega$ is the old momentum/frequency (note that k only lies in the smaller momentum region $<$) and $k'/\omega'$ are new ones. Then perform change of variables in $S_{\text{eff}}$ in Eq.~\ref{new_v}. One would observe that the  $S_0$ is invariant under this re-scale process ( or more precisely the non-interacting parameter is invariant). Other coupling constant shall be also expressed in terms of this new set of variables/fields. Below we handle the theory perturbatively.
 \subsection{First order in interaction}
The first order perturbation involves the average,
\begin{eqnarray}
\label{first_order}
\langle S_{\text{int}} \rangle_{0>} &=&\frac{1}{2!2!}\prod_{i=1,2,3,4}\left[
\int_{|k|<\Lambda } \frac{dk_i}{(2\pi)}\int_{-\infty}^{+\infty} \frac{d\omega_i}{2\pi}
\right]u(4321) \langle \bar\psi_{i_4}(4) \bar \psi_{i_3}(3)  \psi_{i_2}(2) \psi_{i_1} (1)\rangle_{0>}\nonumber \\
&\times&  2\pi \delta(k_4+k_3-k_2-k_1)2\pi \delta(\omega_4+\omega_3-\omega_2-\omega_1)
\end{eqnarray}
Notice that the interaction in the action takes an extra sign. Therefore, $u_{+--+}=-u_0$. Express 4-point correlation into fast and slow modes,
\begin{eqnarray}
\langle \bar\psi_{i_4} \bar \psi_{i_3}  \psi_{i_2} \psi_{i_1} \rangle_{0>} = \langle  (\bar\psi_{i_4,<}+\bar\psi_{i_4,>} ) (\bar\psi_{i_3,<}+\bar\psi_{i_3,>} ) ( \psi_{i_2,<}+ \psi_{i_2,>} )  ( \psi_{i_1,<}+ \psi_{i_1,>} ) \rangle_{0>}
\end{eqnarray}
One can expand the brackets. Non-zero contributions consist of three types: (1). $4$ slow terms, namely, tree-level scaling. (2) $2$ slow /$2$ fast terms. This part gives corrections to Gaussian part of theory. (3) $4$ fast terms, giving a constant in effective action thus negligible. As mentioned in the main-text, one may write the potential as 
\begin{eqnarray}
u_{+-+-}(k_4k_3k_2k_1)\simeq u_{0}+ u_{1}(k_2-k_3) + \text{irrelevant}.
\end{eqnarray}
\begin{itemize}
    \item {\it Tree-level.} This level is nothing but performing power counting based on Eq.~\ref{new_v}. This leads to
\begin{eqnarray}
u'(4'3'2'1')= s^{-3(z+1)+2(2z+1)}u(432 1)=s^{z-1}u(432 1).
\end{eqnarray}
The first part is from the re-scale of variables while the second part is from field operators. Here we consider $z=2$, then$u_0$ becomes a relevant perturbation while $u_0$ is a marginal perturbation.
 
 \item {\it Correction to Gaussian action: $u_0$-part.}  Here we consider the symmetric potential. Via some simple combinatorics, one would see terms like 
\begin{eqnarray}
  -4 u_{+-+-} \bar{\psi}_{<,-}(4) \psi_{<,-}(3)\langle   \bar{\psi}_{>,+} (2) \psi_{>,+} (1) \rangle -4 u_{+-+-}\bar{\psi}_{<,+}(4) \psi_{<,+}(3) \langle   \bar{\psi}_{>,-}(2) \psi_{>,-}(1) \rangle.
\end{eqnarray}
The extra minus sign comes from that we commute two fermion operators. Note that $1$ is a short notation including both momentum and frequency. The average here is simply the Guassian type of integral with
\begin{eqnarray}
\langle \bar{\psi}_{\alpha >} (\omega k) \psi_{\beta >}(\omega'  k') \rangle=\frac{\delta_{\alpha\beta}}{i\omega-\alpha uk^2}  2\pi \delta(k-k') 2\pi \delta(\omega-\omega')
\end{eqnarray}
Thus, in the Eq.~\ref{first_order}, this delta function only preserve one integral in fast modes. The conservation of total momentum make momenta of two slow modes equal. 
\begin{eqnarray}
\langle S_{\text{int}} \rangle_{0>} &=&  
-u_0\int_{|k|<\Lambda/s } \frac{dk}{(2\pi)}\int_{-\infty}^{+\infty} \frac{d\omega}{2\pi}\sum_{\alpha=\pm}
  \bar{\psi}_{<,\alpha}(k\omega)\psi_{<,\alpha}(k\omega) \int_{|q|>\Lambda/s} \frac{dq}{2\pi} \int  \frac{d\omega_1}{2\pi}  \frac{e^{i\omega^+ 0^+}}{i\omega_1+\alpha q^2} \nonumber
\end{eqnarray}
The integral can be explicitly down, e.g., 
\begin{eqnarray}
\int_{|q|>\Lambda/s} \frac{dq}{2\pi} \int  \frac{d\omega_1}{2\pi}  \frac{e^{i\omega_1 0^+}}{i\omega_1+ q^2} =\int_{\Lambda>|q|>\Lambda/s} \frac{dq}{2\pi}= \frac{\Lambda(1-s^{-1})}{\pi}
\end{eqnarray}
Here the integral in the tadpole diagram needs to carefully treated, since the integral over frequency does not converge. The result depends on the regularization condition $e^{i\omega_1 0^+}$. Here we match $G_-(k)$ with  the particle number at $k$-mode in the Dirac Sea and $G_+(k)$ with zero. Therefore, the correction to the Gaussian action reads
\begin{eqnarray}
\langle S_{\text{int}} \rangle_{0>} &=&
-u_0\int_{|k|<\Lambda/s } \frac{dk}{(2\pi)}\int_{-\infty}^{+\infty} \frac{d\omega}{2\pi}\sum_{\alpha=+} 
  \bar{\psi}_{<,\alpha}  (k\omega)\psi_{<,\alpha}(k\omega) \frac{\Lambda(1-s^{-1})}{\pi}.
\end{eqnarray}
 One may add counter terms back to bare Lagarangian so that
RG transformation would not generate
the tadpole contribution.   When the external legs have zero momentum and frequency input, one cannot tell difference between $\psi_+$ and $\psi_-$ anymore. Therefore, it is amounting to a constant shift to the theory.

\item {\it Correction to Gaussian action: $u_1$-part.} . Similarly, one may write the correction from $u_1$ reads
\begin{eqnarray}
-4u_1 (k_4-k_1)  \bar{\psi}_{<,-}(4) \psi_{<,-}(3)\langle   \bar{\psi}_{>,+} (2) \psi_{>,+} (1) \rangle-4u_1(k_4-k_1) \langle\bar{\psi}_{>,-}(4) \psi_{>,-}(3)\rangle  \bar{\psi}_{<,+} (2) \psi_{<,+} (1)  
\end{eqnarray}
Repeating the same procedure in $u_0$, one may find the quadratic correction,
\begin{eqnarray}
\langle S_{\text{int}} \rangle_{0>} &=& u_1
 \int_{|k|<\Lambda/s } \frac{dk}{(2\pi)}\int_{-\infty}^{+\infty} \frac{d\omega}{2\pi}\sum_{\alpha=\pm}
  \bar{\psi}_{<,\alpha}(k\omega)\psi_{<,\alpha}(k\omega) \int_{|q|>\Lambda/s} \frac{dq}{2\pi} \int  \frac{d\omega_1}{2\pi}  \frac{e^{i\omega_1 0^+}}{i\omega_1+\alpha q^2}  \alpha (k-q)\nonumber
\end{eqnarray}
The second term from $\int qdq$ obviously vanishes. Therefore, we are left with 
\begin{eqnarray}
\langle S_{\text{int}} \rangle_{0>} &=&
u_1\int_{|k|<\Lambda/s } \frac{dk}{(2\pi)}\int_{-\infty}^{+\infty} \frac{d\omega}{2\pi}\sum_{\alpha=+} 
  \bar{\psi}_{<,\alpha}  (k\omega)\psi_{<,\alpha}(k\omega) \frac{\Lambda(1-s^{-1})}{\pi}.
\end{eqnarray}
Equivalently, we reach
\begin{eqnarray}
\frac{dv}{dl}=v+ \frac{u_1 \Lambda}{\pi}
\end{eqnarray}
\end{itemize}

\subsection{Second-order in interaction}
Second order perturbation obviously involves the term below,
\begin{eqnarray}
  u_{i_4 i_3 i_2 i_1}(4321) (\bar\psi_{i_4,<}+\bar\psi_{i_4,>} ) (\bar\psi_{i_3,<}+\bar\psi_{i_3,>} ) ( \psi_{i_2,<}+ \psi_{i_2,>} )  ( \psi_{i_1,<}+ \psi_{i_1,>} ) \nonumber\\
  \times u_{i'_4 i'_3 i'_2 i'_1}(4'3'2'1') (\bar\psi_{i'_4,<}+\bar\psi_{i'_4,>} ) (\bar\psi_{i'_3,<}+\bar\psi_{i'_3,>} ) ( \psi_{i'_2,<}+ \psi_{i'_2,>} )  ( \psi_{i'_1,<}+ \psi_{i'_1,>} )
\end{eqnarray}
Totally, there are $2^8$ terms. Since the average is done over the Gaussian distribution, one only needs to trace even terms.
\begin{itemize}
    \item $8 \psi_>/0 \psi_< $. This term is purely giving constant contribution.
    
    \item $6 \psi_>/2 \psi_<$. Gives renormalization to the non-interacting part but higher order in $u_0$.
    
    \item $4 \psi_>/4 \psi_< $. This term renormalizes the quartic interaction.
    
    \item $2\psi_>/6\psi_<$, $0\psi_>/8\psi_<$. These terms are indeed generated in the re-normalization process. 
\end{itemize}
 {\it  Corrections to the quartic term. }  
To list all possible terms generating quartic corrections, one may enumerate all possible terms in some principle. Since two interaction potentials are involved, one may list terms from (1). Two come-in (creation) operators from one potential while two come-out (annihilation) operators arise from the one potential. (2) One come-in is from one potential while the other come-in is from the one potential. From this principle, there is only two types. 
To symmetrize the local potential in momentum space, the second type is classified into two types. To see how this happen, let us consider the short-hand representation below ( which is used to analyze the expansion),
\begin{eqnarray}
\label{23}
 u(4321) \bar\psi(4)  \bar\psi(3)  \psi(2)\psi(1) \times  u(8765) \bar\psi(8)  \bar\psi(7)  \psi(6)\psi(5).
\end{eqnarray}
Each field operator here contains both $>$ and $<$. Now consider one piece of contribution in Eq.~\ref{23} 
 \begin{eqnarray}
 u(8765) u(4321)\,\bar\psi_>(4)\,  \bar\psi_<(3)\,  \psi_>(2)\,\psi_<(1)    \, \bar\psi_>(8) \, \bar\psi_<(7) \, \psi_>(6)\, \psi_<(5).
\end{eqnarray}
Note that the average will be done on the $>$ components. Thus a simple re-arrangement leads to 
 \begin{eqnarray}
 -u(8765)  u(4321)  \bar\psi_<(3) \,   \bar\psi_<(7)\,  \psi_<(1) \psi_<(5)\, 
 \langle 
 \bar\psi_>(4) \,  \psi_>(2)\,  \bar\psi_>(8)\,  \psi_>(6)  \, 
 \rangle
\end{eqnarray}
There is only one way to pair field operators in a connected way,
\begin{eqnarray}
 u(8765)  u(4321)  \bar\psi_<(3) \,   \bar\psi_<(7)\,  \psi_<(1) \psi_<(5)\, 
 G(4)\delta(4,6)  G(2)\delta(2,8)  
\end{eqnarray}
Then taking integral and delta-function (momentum/energy conservations) into consideration, one finds that $S_{\text{int}}^2$ contains the contribution like
\begin{eqnarray}
\label{ttt}
\delta(4+3-2-1) \bar\psi_<(4) \,   \bar\psi_<(3)\,  \psi_<(2) \psi_<(1)\, \int d5d6
 G(5)  G(6)  \delta (3+5-1-6) u(6452)u(5361)
\end{eqnarray}
In fact, exchanging the role of $1$ and $2$ leads to another form of expression,
\begin{eqnarray}
-\delta(4+3-2-1) \bar\psi_<(4) \,   \bar\psi_<(3)\,  \psi_<(2) \psi_<(1)\, \int d5d6
 G(5)  G(6)  \delta (3+5-2-6) u(6451)u(5362)\nonumber
\end{eqnarray}
This exchange seems to be trivial but important in obtaining a symmetric vertex correction. Since there is totally $2^4$ terms in second type contributing exactly same as Eq.~\ref{ttt}, one may split them into so-called $ZS$ and $ZS'$ diagrams. To be consistent with the reference, we play some some variable change and reach,
\begin{eqnarray}
ZS= \bar\psi_<(4) \,   \bar\psi_<(3)\,  \psi_<(2) \psi_<(1)\, \int d5d6
 G(5)  G(6)  \delta (3+5-1-6) u(6351)u(4526)\\
 ZS'=-\bar\psi_<(4) \,   \bar\psi_<(3)\,  \psi_<(2) \psi_<(1)\, \int d5d6
 G(5)  G(6)  \delta (3+5-2-6) u(6451)u(3526)  
\end{eqnarray}
Here the identity coefficient comes from $8/(2\times 2!\times 2!)=1$. Of course, the first type of contribution is so-called $BCS$ diagram. It contributes as 
\begin{eqnarray}
BCS=-\frac{1}{2}\bar\psi_<(4) \,   \bar\psi_<(3)\,  \psi_<(2) \psi_<(1)\, \int d5d6
 G(5)  G(6)  \delta (5+6-4-3) u(4365)u(6521) \nonumber
\end{eqnarray}
The coefficient $1/2$ comes from $4/8$. Now we have symbolically represented one-loop level contributions. One would see that this is equivalent to usual Feynman diagrams of four-point correlations. Now we consider the loop-correction to $u(+-+-)$, i.e., $4=+,3=-,2=+,1=-$.  Notice that this sector contains two coupling constants,
\begin{eqnarray}
u(+-+-)(k_4k_3k_2k_1)=u_0+u_1(k_2-k_3)
\end{eqnarray}
\begin{itemize}
    \item {\it ZS diagram.} From the specific potential condition, we demand that the sub-indices part satisfies
    \begin{eqnarray}
    63=51=54=26=\{+,-\}.
    \end{eqnarray}
 It is easy to see that this is impossible. Thus contribution from ZS diagram to this typical potential vanishes.
    
     \item {\it ZS' diagram.} From the specific potential condition, we demand that the sub-indices part satisfies
    \begin{eqnarray}
    64=51=35=26=\{+,-\}.
    \end{eqnarray}
    It easy to conclude that $6=-,5=+$. The integral involved reads
    \begin{eqnarray}
 ZS'= -2\int_{\Lambda/s}^{\Lambda} \frac{d k_5}{2\pi} \int_{-\infty}^{+\infty} \frac{d\omega}{2\pi}  G_+(i\omega,k_5)  G_-(i\omega,k_6=k_5-q)    u(-++-)(k_6k_4k_5k_1)u(-++-) (k_3k_5k_2k_6).
    \end{eqnarray} 
Expand two potentials,    
    \begin{eqnarray}  
  u(-++-)(k_6k_4k_5k_1)=-u_0-u_1 (k_5-k_6),\quad u(-++-) (k_3k_5k_2k_6)=-u_0-u_1 (k_2-k_3)
    \end{eqnarray}
    where $q$ is the net come-in momentum. Notice that $q=k_5-k_6=k_2-k_3$ in $ZS'$ channel. Here the factor 2 is due to that the integral in $k<0$ and $k>0$ gives the same contribution. The integral is given by
    \begin{eqnarray}
 ZS'=-2\int_{\Lambda/s}^{\Lambda} \frac{dk}{2\pi} \int_{-\infty}^{+\infty}\frac{e^{i0^+\omega}d\omega}{2\pi}  \frac{ 1}{i\omega- k^2} \frac{1}{i\omega+   (k-q)^2} (u_0+u_1 q)^2
 \end{eqnarray}
The $u^2_0$ contribution gives 
\begin{eqnarray}
 ZS'\Big|_{u_0^2}=-2u_0^2\int_{\Lambda/s}^{\Lambda} \frac{dk}{2\pi} \int_{-\infty}^{+\infty}\frac{e^{i0^+\omega}d\omega}{2\pi}  \frac{ 1}{i\omega- k^2} \frac{1}{i\omega+   (k-q)^2}=2u_0^2\int_{\Lambda/s}^{\Lambda} \frac{dk}{2\pi}    \frac{1}{k^2+   (k-q)^2}.
 \end{eqnarray}
The integral on the shell is given by
\begin{eqnarray}
 ZS'\Big|_{u_0^2}= \frac{u_0^2 d\Lambda}{2\pi \Lambda^2 }\frac{1}{1-2q/\Lambda+q^2/\Lambda^2}
 \simeq \frac{u_0^2 d\Lambda}{2\pi \Lambda^2 } \left(1+\frac{2q}{\Lambda} +O(q^2)\right).
 \end{eqnarray}
This piece give contributions to the change of each coupling constant, $ 
du_0=\frac{u_0^2 d\Lambda}{2\pi \Lambda^2 },\quad du_1=\frac{u_0^2 d\Lambda}{ \pi \Lambda^3 }.
$  Then let us consider $u_0 u_1$ contribution, 
\begin{eqnarray}
 ZS'\Big|_{u_0 u_1}=-4 u_0u_1 q \int_{\Lambda/s}^{\Lambda} \frac{dk}{2\pi} \int_{-\infty}^{+\infty}\frac{e^{i0^+\omega}d\omega}{2\pi}  \frac{ 1}{i\omega- k^2} \frac{1}{i\omega+   (k-q)^2} = 2 u_0u_1 \frac{  d\Lambda}{2\pi \Lambda^2 } q
 \end{eqnarray}
Therefore, one may reach the RG equation for each couple constant,
\begin{eqnarray}
\frac{du_0}{dl}=u_0 + \frac{u_0^2 }{2\pi\Lambda },
\\
\frac{du_1}{dl}=\frac{u_0^2 }{ \pi \Lambda^2 }+\frac{ u_0u_1 }{\pi \Lambda }.
\end{eqnarray}

\item {\it BCS diagram.} Now consider BCS contribution.  We demand that the sub-indices part satisfies  $\{5,6\}=\{+,-\}$. There are two options. They contribute,
\begin{eqnarray}
BCS&=&-\frac{u^2}{2}\int_{\Lambda/s}^{\Lambda} \frac{dk}{2\pi} \int_{-\infty}^{+\infty} \frac{d\omega}{2\pi}  G_+(i\omega,k)  G_-(i\nu-i\omega,q-k) \nonumber \\
&-&\frac{u^2}{2}\int_{\Lambda/s}^{\Lambda} \frac{dk}{2\pi} \int_{-\infty}^{+\infty} \frac{d\omega}{2\pi}  G_-(i\omega,k)  G_+(i\nu-i\omega,q-k)
\end{eqnarray}
Still trace the zero input frequency and momentum,
\begin{eqnarray}
BCS&=&\frac{u^2}{2}\int_{\Lambda/s}^{\Lambda} \frac{dk}{2\pi} \int_{-\infty}^{+\infty}\frac{d\omega}{2\pi}  \left(
\frac{1}{i\omega-   k^2} \frac{1}{i\omega-   k^2}+\frac{1}{i\omega+   k^2} \frac{1}{i\omega+   k^2}
\right)
\end{eqnarray}
Note that the residue of $1/z^2$ is zero. Thus the BCS contribution is zero.
\end{itemize}

 \subsection{Third order in Interaction}
We have derived the RG equation at the loop level. Also we only have {
\it one
} running coupling. Now natural questions arises: is one-loop RG reliable?  We answer these questions in the scheme of the momentum shell RG. Below we use third order perturbations to argue why we can stop at one-loop RG.

 {\bf Third order perturbations}
The third order perturbation involves the term below,
\begin{eqnarray}
\label{64}
  u_{i_4 i_3 i_2 i_1}(4321) (\bar\psi_{i_4,<}+\bar\psi_{i_4,>} ) (\bar\psi_{i_3,<}+\bar\psi_{i_3,>} ) ( \psi_{i_2,<}+ \psi_{i_2,>} )  ( \psi_{i_1,<}+ \psi_{i_1,>} ) \nonumber\\
  \times u_{i'_4 i'_3 i'_2 i'_1}(4'3'2'1') (\bar\psi_{i'_4,<}+\bar\psi_{i'_4,>} ) (\bar\psi_{i'_3,<}+\bar\psi_{i'_3,>} ) ( \psi_{i'_2,<}+ \psi_{i'_2,>} )  ( \psi_{i'_1,<}+ \psi_{i'_1,>} )\nonumber\\
  \times u_{i''_4 i''_3 i''_2 i''_1}(4''3''2''1'') (\bar\psi_{i''_4,<}+\bar\psi_{i''_4,>} ) (\bar\psi_{i''_3,<}+\bar\psi_{i''_3,>} ) ( \psi_{i''_2,<}+ \psi_{i''_2,>} )  ( \psi_{i''_1,<}+ \psi_{i''_1,>} )
\end{eqnarray}
Totally, there are $2^{12}$ terms. Still, one only has to consider terms with even power of operators in $<$. Let us go through all possible terms
\begin{itemize}
    \item $12 \psi_>/0 \psi_< $. It is purely giving constant contribution.
    
    \item $10 \psi_>/2 \psi_<$. It gives renormalization to the non-interacting part but in third order of $u_0$.  
    
    \item $8 \psi_>/4 \psi_< $. Higher order of renormalizing the quartic interaction, which is  basically the  two-loop correction.  In the momentum shell RG, this term contributes $\propto (d\Lambda) ^2$. In the linear equation, this term does not contribute to the running coupling at all. One has to consider the higher order contribution from high order vertices. See below.
    
    \item $6\psi_>/6\psi_<$. This term would give six-point vertex, which is important if one wants to go beyond the second order perturbation.  
    
    \item $4\psi_>/8\psi_<$, $2\psi_>/10\psi_<$, $0\psi_>/12\psi_<$. Multi-point vertex.
     
\end{itemize}
{\it Six-point Vertex: a new interaction.} Before doing the technical calculations, let us write down the general form of 3-body interactions ( which is basically a 6-point vertex ),
\begin{eqnarray}
H_{3}= 
\prod_{i=1,..,6}\left[
\int_{|k|<\Lambda } \frac{dk_i}{(2\pi)}\int_{-\infty}^{+\infty} \frac{d\omega_i}{2\pi}
\right]g_{i_6 i_5i_4 i_3 i_2 i_1}(654321)  \bar\psi_{i_6}(6) \bar\psi_{i_5}(5) \bar \psi_{i_4}(4) \psi_{i_3}(3) \psi_{i_2}(2) \psi_{i_1} (1) \nonumber \\
 \times   2\pi \delta(k_6+k_5+k_4-k_3-k_2-k_1)2\pi \delta(\omega_6+\omega_5+\omega_4-\omega_3-\omega_2-\omega_1).
\end{eqnarray}
A symmetrized representation of the potential  demands that exchanging any two indices in $6,5,4$ or $3,2,1$ introduces a negative sign to $g$, e.g., $g_{i_6 i_5i_4 i_3 i_2 i_1}(654321) =-g_{i_5i_6  i_4 i_3 i_2 i_1}(564321)$. It is basically described by two distinct elements,
\begin{eqnarray}
g_p\equiv g_{++-;-++}(k_6,k_5,k_4;k_3,k_2,k_1),\quad g_n\equiv g_{+--;--+}(k_6,k_5,k_4;k_3,k_2,k_1).
\end{eqnarray}
Here $+/-$ denotes particle/hole species: $g_p$ describes $2$ particles and $1$ hole collides, while $g_n$ describes $2$ holes and $1$ particles collides. Generally other types of interaction may also exist. But if we focus the situation where $g$-interaction is generated by $u$-interaction, these two are only allowed terms.

\begin{figure}[h]
\includegraphics[width=4in,clip]{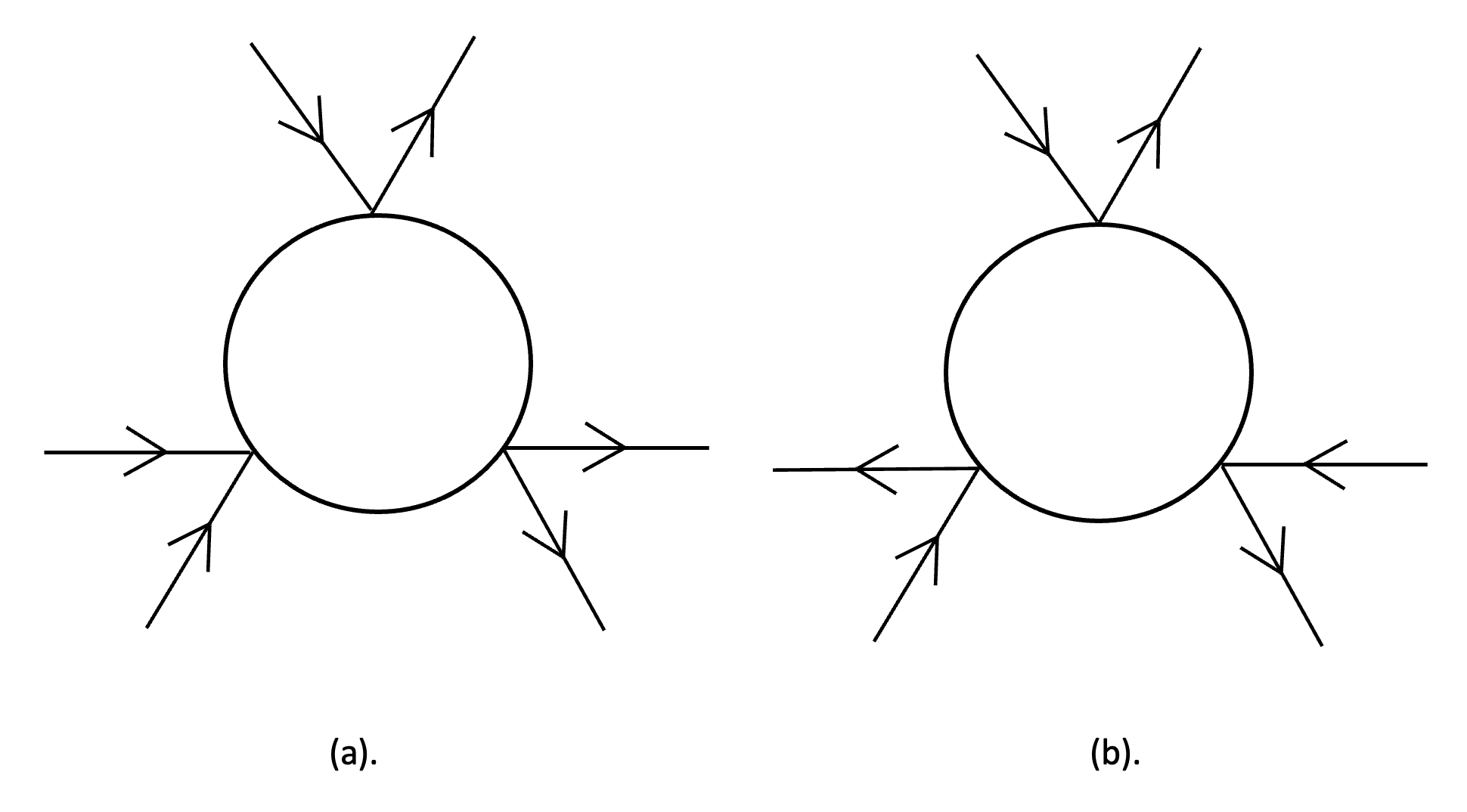}
  \caption{(Color online) Two types of diagrams contributing to 6-point vertex.  }
  \label{tensor}
\end{figure}

{\it Generated Diagrams. }To make discussion more complete, we consider how $H_3$($g$-interaction) is generated by $u$-interaction. There are two points to be emphasized. 
\begin{itemize}
    \item Firstly we only consider the contributions from the connected Feynman diagrams. This condition excludes the possibility of four $<$ operators (slower modes) coming from a single $u$-vertex (2-body interaction).
    \item  Secondly, we would set all external momentum to be zero in the final step to consider the coupling {\it constant}.  Therefore, the momentum conservation exclude the possibility of  3 $<$ operators coming from a single 4-point vertex.
\end{itemize}
 As a conclusion, one only has to consider such a situation: each 4-point vertex contribute 2 $>$ operators and 2 $<$ operators. Further more, we want the equal number of creation and annihilation operators for either $>$ or $<$ types. One would see two different situations to get $6\psi_>/6\psi_<$:
 \begin{eqnarray}
\text{Case I}:\quad \bar{\psi}_<(12) \bar{\psi}_<(11)   {\psi}_>(10)  {\psi}_>(9) \times   \bar{\psi}_>(8) \bar{\psi}_>(7)   {\psi}_<(6)  {\psi}_<(5) \times  \bar{\psi}_< (4)\bar{\psi}_>(3)   {\psi}_<(2)  {\psi}_>(1). \quad M=3! \times 4 \\
\text{Case II}: \bar{\psi}_<(12) \bar{\psi}_> (11)  {\psi}_<(10)  {\psi}_>(9) \times  \bar{\psi}_<(8) \bar{\psi}_>(7)   {\psi}_< (6) {\psi}_>(5) \times  \bar{\psi}_<(4) \bar{\psi}_>(3)   {\psi}_<(2)  {\psi}_>(1) .\quad M=2^6.
 \end{eqnarray}
 Here each $4$ field operator is assumed to belonging to a single $u$-vertex. The $M$ denotes the multiplicity for each the case. The Case I corresponds to the Diagram in Fig.~\ref{tensor}a. Scattering events in three vertices are described by particle-particle, particle-particle and particle-hole scatterings. The case II corresponds to the diagram in Fig.~\ref{tensor}b. Scattering events in three vertices are described by particle-hole, particle-hole and particle-hole scatterings. 

The next step would be contracting the $<$-operators. Now we consider two cases separately. 
\begin{itemize}
    \item {\it Case I.} One has to consider the term 
    \begin{eqnarray}
    \label{68}
  -u(12,11,10,9)  u(8765)   u(4321)    \bar{\psi}_<(12) \bar{\psi}_<(11)        {\psi}_<(6)  {\psi}_<(5)   \bar{\psi}_< (4)    {\psi}_<(2)  \nonumber\\
 \times       \langle {\psi}_>(10)  {\psi}_>(9) \bar{\psi}_>(8) \bar{\psi}_>(7) \bar{\psi}_>(3) {\psi}_>(1) \rangle
    \end{eqnarray}
    Here the negative sign comes from moving all $>$ operator in the front of average. There are totally 4 ways to do the contraction.
    \begin{eqnarray}
     \langle {\psi}_>(10)  {\psi}_>(9) \bar{\psi}_>(8) \bar{\psi}_>(7) \bar{\psi}_>(3) {\psi}_>(1) \rangle=-G(8)\delta(10,8)G(3)\delta(9,3) G(1)\delta(7,1)+G(7)\delta(10,7)G(3)\delta(9,3) G(1)\delta(8,1) \nonumber\\
     +G(8)\delta(9,8)G(3)\delta(10,3) G(1)\delta(7,1)-G(7)\delta(9,7)G(3)\delta(10,3) G(1)\delta(8,1) \nonumber
    \end{eqnarray}
   The relation between first and second term in the first line is exchanging the index $7$ and $8$. The  relation between the first line and second line is exchanging the role of $9$ and $10$. Because the anti-symmetric property of the potential $u(4321)$, four terms in this contraction give exactly same contribution. Namely, Case I contribute to $H3$ with the form,
 \begin{eqnarray}
\frac{3!\times 4\times 4}{(2)^6}u(12,11,8,3)  u(8165)   u(4321)    \bar{\psi}_<(12) \bar{\psi}_<(11)  \bar{\psi}_< (4)       {\psi}_<(6)  {\psi}_<(5)       {\psi}_<(2)    G(8) G(3) G(1). 
\end{eqnarray}
Re-labeling the indices, one obtains 
\begin{eqnarray}
\frac{3!\times 4\times 4}{3!(2)^6}   \int d1' d2' d3' u(65 1' 2')  u(1'3'32)   u(42'13')    \bar{\psi}_<(6) \bar{\psi}_<(5)  \bar{\psi}_< (4)       {\psi}_<(3)  {\psi}_<(2)       {\psi}_<(1)    G(1') G(2') G(3'). 
\end{eqnarray}
Here $1', 2'$ and $3'$ are three internal variables. Upon summation all possible configurations of six ${123456}$ indices, one may obtain the Case. I. contribution to the 6-point vertex. But this is still too complicated. Similar to the situation of 4-point vertex, one may derive a symmetrized version of 6-point vertex.  
\item {\it Case II.} One needs to consider the term
 \begin{eqnarray}
 \label{70}
   u(12,11,10,9)  u(8765)   u(4321)   \bar{\psi}_<(12)\bar{\psi}_<(8) \bar{\psi}_<(4) {\psi}_<(10) {\psi}_< (6)   {\psi}_<(2)\nonumber \\
   \langle \bar{\psi}_> (11)     {\psi}_>(9)    \bar{\psi}_>(7)     {\psi}_>(5)   \bar{\psi}_>(3)     {\psi}_>(1)\rangle
\end{eqnarray}
There are only two ways to perform contraction,
 \begin{eqnarray}
 \langle \bar{\psi}_> (11)     {\psi}_>(9)    \bar{\psi}_>(7)     {\psi}_>(5)   \bar{\psi}_>(3)     {\psi}_>(1)\rangle=-G(5) \delta(5,11)G(3) \delta(3,9)G(1) \delta(1,7)-G(1) \delta(1,11)G(7) \delta(7,9)G(3) \delta(3,5) \nonumber
 \end{eqnarray}
 Notice that two contributions are intrinsically different. Therefore Eq.~\ref{70} becomes
\begin{eqnarray}
   -u(12,5,10,3)  u(8165)   u(4321)   \bar{\psi}_<(12)\bar{\psi}_<(8) \bar{\psi}_<(4) {\psi}_<(10) {\psi}_< (6)   {\psi}_<(2)G(5)G(3)G(1) \nonumber \\
  -u(12,1,10,7)  u(8763)   u(4321)   \bar{\psi}_<(12)\bar{\psi}_<(8) \bar{\psi}_<(4) {\psi}_<(10) {\psi}_< (6)   {\psi}_<(2)  G(1)  G(7)  G(3)  
\end{eqnarray}
Re-labeling the indices, one obtains
 \begin{eqnarray}
   -u(6,1',3,2')  u(53'21')   u(42'13')   \bar{\psi}_<(6)\bar{\psi}_<(5) \bar{\psi}_<(4) {\psi}_<(3) {\psi}_< (2)   {\psi}_<(1)G(1')G(2')G(3') \nonumber \\
  -u(6,1',3,2')  u(52'23')   u(43'11')   \bar{\psi}_<(6)\bar{\psi}_<(5) \bar{\psi}_<(4) {\psi}_<(3) {\psi}_< (2)   {\psi}_<(1)  G(1')  G(2')  G(3')  
\end{eqnarray}
\end{itemize}

{\it No coupling Constant from $g$-interaction.}
Since the underlying degrees of freedom only carry two internal species $+$ and $-$, there would be no space for coulping constants $g_p$ and $g_n$. For example,
\begin{eqnarray}
g_{p}(k_6,k_5,k_4;k_3,k_2,k_1)=-g_{p}(k_5,k_6,k_4;k_3,k_2,k_1)
\end{eqnarray}
If we set all external momentum to be zero, as we always do to obtain a coupling constant, one immediately reach
\begin{eqnarray}
g_{p}(0)=-g_{p}(0) \rightarrow g_{p}(0)=0. 
\end{eqnarray}
Therefore, the coupling constant part of $g-$interaction vanishes. Of course one can assume it to be a {\it coupling function }. However, via expanding the function in the small momentum transfer, one would see all expansions are irrelevant perturbations from the power counting. In this sense, one can ignore the contribution from $g$-interactions.  This consideration may justify why we can stop at the one-loop level. 

\end{document}